\documentclass[showpacs,preprintnumbers,amsmath,amssymb]{revtex4}

\usepackage{epsfig}
\usepackage{graphicx}
\usepackage{dcolumn}
\usepackage{bm}

\begin{document}


\title{Charge and Spin Quantum Fluids Generated by Many-Electron Interactions}
\author{J. M. P. Carmelo}
\affiliation{GCEP-Center of Physics, University of Minho, Campus
Gualtar, P-4710-057 Braga, Portugal}
\author{J. M. Rom\'an}
\affiliation{GCEP-Center of Physics, University of Minho, Campus
Gualtar, P-4710-057 Braga, Portugal}
\author{K. Penc}
\affiliation{Research Institute for Solid State Physics and
Optics, H-1525 Budapest, P.O.B. 49, Hungary}
\date{21 November 2003}


\begin{abstract}
In this paper we describe the electrons of the non-perturbative
one-dimensional (1D) Hubbard model by a fluid of unpaired rotated
electrons and a fluid of zero-spin rotated-electron pairs. The
rotated electrons are related to the original electrons by a mere
unitary transformation. For {\it all} finite values of energy and
for the whole parameter space of the model this two-fluid picture
leads to a description of the energy eigenstates in terms of
occupancy configurations of $\eta$-spin $1/2$ holons, spin $1/2$
spinons, and $c$ pseudoparticles only. The electronic degrees of
freedom couple to external charge (and spin) probes through the
holons and $c$ pseudoparticles (and spinons). Our results refer to
very large values of the number of lattice sites $N_a$. The holon
(and spinon) charge (and spin) transport is made by $2\nu$-holon
(and $2\nu$-spinon) composite pseudoparticles such that
$\nu=1,2,...$. For electronic numbers obeying the inequalities
$N\leq N_a$ and $N_{\downarrow}\leq N_{\uparrow}$ there are no
zero-spin rotated-electron pairs in the ground state and the
unpaired-rotated-electron fluid is described by a charge $c$
pseudoparticle fluid and a spin $\nu=1$ two-spinon pseudoparticle
fluid. The spin two-spinon pseudoparticle fluid is the 1D
realization of the two-dimensional {\it resonating valence bond}
spin fluid.
\end{abstract}

\pacs{70}

\maketitle
\section{INTRODUCTION}

The relation for the whole Hilbert space of the non-perturbative
one-dimensional (1D) Hubbard model \cite{Lieb,Takahashi} of the
original electrons to the quantum objects whose occupancy
configurations describe its energy eigenstates is an interesting
and important open problem. Such a non-perturbative relation is
needed for the study of the finite-energy few-electron spectral
properties of the many-electron quantum problem. The main goal of
this paper is the study of such a relation. Our study is motivated
by both the unusual finite-energy spectral properties observed in
quasi-1D materials \cite{spectral0,Gweon} and the relation of
these materials to two-dimensional (2D) quantum problems
\cite{Granath,Zaanen,Antonio}.

In a chemical potential $\mu $ and magnetic field $h$ the 1D
Hubbard Hamiltonian can be written as,

\begin{equation}
\hat{H}={\hat{H}}_{SO(4)} + \sum_{\alpha =c,\,s}\mu_{\alpha
}\,2{\hat{S}}_{\alpha}^z \, , \label{H}
\end{equation}
where $\mu_c=\mu$ is the chemical potential, $\mu_s=\mu_0 h$,
$\mu_0$ is the Bohr magneton, and the number operators,

\begin{equation}
{\hat{S }}_c^z=-{1\over 2}[N_a-\hat{N}] \, ; \hspace{1cm} {\hat{S
}}_s^z= -{1\over 2}[{\hat{N}}_{\uparrow}- {\hat{N}}_{\downarrow}]
\, , \label{Sz}
\end{equation}
are the diagonal generators of the $\eta$-spin and spin $SU(2)$
algebras \cite{HL,Yang89}, respectively. The Hamiltonian
${\hat{H}}_{SO(4)}$ on the right-hand side of Eq. (\ref{H}) reads,

\begin{equation}
{\hat{H}}_{SO(4)} = {\hat{H}}_H  - (U/2)\,\hat{N} + (U/4)\,N_a \,
, \label{HH}
\end{equation}
where

\begin{equation}
{{\hat{H}}}_H = \hat{T} + U\,\hat{D} \, , \label{HHS}
\end{equation}
is the Hubbard model in standard notation. In the latter
expression,

\begin{equation}
\hat{T} = -t\sum_{j =1}^{N_a}\sum_{\sigma =\uparrow
,\downarrow}\sum_{\delta =-1,+1}
c_{j,\,\sigma}^{\dag}\,c_{j+\delta,\,\sigma} \, , \label{opT}
\end{equation}
is the {\it kinetic-energy} operator and

\begin{equation}
\hat{D} = \sum_{j}c_{j,\,\uparrow}^{\dag}\,c_{j,\,\uparrow}
c_{j,\,\downarrow}^{\dag}\,c_{j,\,\downarrow} =
\sum_{j}\hat{n}_{j,\,\uparrow}\,\hat{n}_{j,\,\downarrow}   \, ,
\label{Dop}
\end{equation}
is the electron double-occupation operator. We call the kinetic
energy $T$ and number of electron doubly occupied sites $D$ the
expectations values of the operators (\ref{opT}) and (\ref{Dop}),
respectively.

The Hamiltonian given in Eq. (\ref{HH}) has $SO(4)$ symmetry
\cite{HL,Yang89,II} and commutes with the six generators of the
$\eta$-spin and spin algebras. The expressions of the two
corresponding diagonal generators are given in Eq. (\ref{Sz}) and
the off-diagonal generators of these two $SU(2)$ algebras read,

\begin{equation}
{\hat{S}}_c^{\dagger}=\sum_{j}(-1)^j\,c_{j,\,\downarrow}^{\dagger}\,
c_{j,\,\uparrow}^{\dagger} \, ; \hspace{1cm} {\hat{S}}_c
=\sum_{j}(-1)^j\,c_{j,\,\uparrow}\,c_{j,\,\downarrow} \, ,
\label{Sc}
\end{equation}
and

\begin{equation}
{\hat{S}}_s^{\dagger}=
\sum_{j}\,c_{j,\,\downarrow}^{\dagger}\,c_{j,\,\uparrow} \, ;
\hspace{1cm} {\hat{S}}_s=\sum_{j}c_{j,\,\uparrow}^{\dagger}\,
c_{j,\,\downarrow} \, , \label{Ss}
\end{equation}
respectively. The $\eta$-spin and spin square operators have
eigenvalue $S_{\alpha}[S_{\alpha}+1]$ with $\alpha =c$ and $\alpha
=s$, where $S_c$ and $S_s$ denote the $\eta$ spin and spin values,
respectively.

We consider that the number of lattice sites $N_a$ is large and
even and $N_a/2$ is odd. The electronic number operators of Eq.
(\ref{Sz}) read ${\hat{N}}=\sum_{\sigma}\hat{N}_{\sigma}$ and
${\hat{N}}_{\sigma}=\sum_{j}\hat{n}_{j,\,\sigma}$ where $
\hat{n}_{j,\,\sigma} = c_{j,\,\sigma }^{\dagger }\,c_{j,\,\sigma
}$. Here $c_{j,\,\sigma}^{\dagger}$ and $c_{j,\,\sigma}$ are spin
$\sigma $ electron operators at site $j=1,2,...,N_a$. We denote
the lattice constant by $a$ and the system length by $L=a\,N_a$.
Throughout this paper we use units of Planck constant one. We
consider electronic densities $n=N/L$ and spin densities
$m=[N_{\uparrow}-N_{\downarrow}]/L$ in the domains $0\leq n \leq
1/a$\, ; $1/a\leq n \leq 2/a$ and $-n\leq m \leq n$\, ;
$-(2/a-n)\leq m \leq (2/a-n)$, respectively.

The Bethe-ansatz solvability of the 1D Hubbard model (\ref{H}) is
restricted to the Hilbert subspace spanned by the lowest-weight
states (LWSs) or highest-weight states (HWSs) of the $\eta$-spin
and spin algebras, {\it i.e.} by the states whose $S_{\alpha}$ and
$S_{\alpha}^z$ numbers are such that $S_{\alpha}= -S_{\alpha}^z$
or $S_{\alpha}=S_{\alpha}^z$, respectively, where $\alpha =c$ for
$\eta$ spin and $\alpha =s$ for spin. In this paper we choose the
$\eta$-spin and spin LWSs description of the Bethe-ansatz
solution. In this case, that solution describes energy eigenstates
associated with densities and spin densities in the domains $0\leq
n \leq 1/a$ and $0\leq m \leq n$, respectively. The description of
the states corresponding to the extended $n$ and $m$ domains
mentioned above is achieved by application onto the latter states
of off-diagonal generators of the $\eta$-spin and spin $SU(2)$
algebras \cite{Essler}. Recently, universal properties of the
model (\ref{H}) entropy were considered \cite{Korepin}.

It is well known that the low-energy eigenstates of the model
(\ref{H}) can be described by occupancy configurations of holons
and spinons \cite{Anderson,PWA,Penc,hs,Tohyama}. On the other
hand, in Ref. \cite{Carmelo97} it was found that all energy
eigenstates associated with the 1D Hubbard model Bethe-ansatz
solution \cite{Lieb,Takahashi} can be described in terms of
occupancy configurations of pseudoparticles. According to the
studies of Ref. \cite{Carmelo97} there is an infinite number of
pseudoparticle branches: the $c$ pseudoparticles and the
$\alpha\nu$ pseudoparticles such that $\alpha =c,\,s$ and
$\nu=1,2,...$. (The $\alpha,\gamma$ pseudoparticle notation of
Ref. \cite{Carmelo97} is such that $\gamma =\nu
-\delta_{\alpha,\,s}$.) The thermodynamic coupled non-linear
equations introduced by Takahashi \cite{Takahashi} can be
understood as describing a Landau liquid of such pseudoparticles.
In Appendix A we relate the the quantum numbers of the latter
equations to the discrete pseudoparticle momentum values. In
Appendix B we provide some aspects of the pseudoparticle
description that are useful for the studies of this paper. The
physical quantities can be expressed as functionals of the
pseudoparticle momentum distribution functions whose occupancies
describe the energy eigenstates. For instance, the $\eta$-spin
value $S_c$, spin value $S_s$, and momentum $P$ read,

\begin{equation}
S_c = {1\over 2}{L\over 2\pi}\Bigl[\int_{q_c^{-}}^{q_c^{+}}\,dq\,
[1-N_c (q)] - 2\sum_{\nu =1}^{\infty} \int_{-q_{c\nu }}^{q_{c\nu
}}\,dq\, \nu N_{c\nu} (q) \Bigr] \, , \label{Scco}
\end{equation}

\begin{equation}
S_s = {1\over 2}{L\over 2\pi}\Bigl[ \int_{q_c^{-}}^{q_c^{+}}\,dq\,
N_c (q) - 2\sum_{\nu =1}^{\infty}
\int_{-q_{s\nu}}^{q_{s\nu}}\,dq\, \nu N_{s\nu} (q)\Bigr] \, ,
\label{Ssco}
\end{equation}
and

\begin{equation}
P = {L\over 2\pi}\Bigl\{\int_{q_c^{-}}^{q_c^{+}}\,dq\, N_c (q)\, q
+ \sum_{\nu =1}^{\infty} \int_{-q_{s\nu }}^{q_{s\nu }}\,dq\,
N_{s\nu} (q)\, q + \sum_{\nu =1}^{\infty}\int_{-q_{c\nu
}}^{q_{c\nu }}\,dq\, N_{c\nu} (q)\, [{\pi\over a}(1+\nu )
-q]\Bigr\} \, , \label{Ppco}
\end{equation}
respectively, where when $\vert\,P\vert >\pi/a$ the value of the
momentum should be brought to the first Brillouin zone. The
limiting pseudoparticle momentum values of these equations are
given in Appendix B. Moreover, the energy spectrum associated with
the Hamiltonian (\ref{HH}) can be written as,

\begin{eqnarray}
E_{SO(4)} & = & -2t {L\over 2\pi} \int_{q_c^{-}}^{q_c^{+}}\,dq\,
N_c (q) \cos (k(q)\,a)
\nonumber \\
& + & 4t {L\over 2\pi}\sum_{\nu=1}^{\infty} \int_{-q_{c\nu
}}^{q_{c\nu}}\,dq\, N_{c\nu} (q)\,{\rm Re}\,\Bigl\{\sqrt{1 -
(\Lambda_{c\nu} (q) + i \nu U/4t)^2}\Bigr\} \nonumber \\
& + & {L\over 2\pi}{U\over 2} \Bigl[\int_{q_c^{-}}^{q_c^{+}}\,dq\,
[{1\over 2}-N_c (q)] - 2\sum_{\nu =1}^{\infty} \int_{-q_{c\nu
}}^{q_{c\nu }}\,dq\, \nu N_{c\nu} (q) \Bigr] \, . \label{Epso4co}
\end{eqnarray}
Here $k(q)$ and $\Lambda_{\alpha\nu} (q)$ with $\alpha =c$ are the
rapidity-momentum functional and the $\alpha\nu$ rapidity
functional, respectively. These functionals are defined by the
Takahashi's thermodynamic equations which can be rewritten in
functional form as follows,

\begin{eqnarray}
k(q) & = & q - {1\over \pi\,a}\sum_{\nu =1}^{\infty}
\int_{-q_{s\nu}}^{q_{s\nu}}\,dq'\, N_{s\nu}(q')\arctan\Bigl({\sin
(k(q)\,a)-\Lambda_{s\nu}(q') \over \nu U/4t}\Bigr)
\nonumber \\
& - & {1\over \pi\,a}\sum_{\nu =1}^{\infty}
\int_{-q_{c\nu}}^{q_{c\nu}}\,dq'\, N_{c\nu}(q') \arctan\Bigl({\sin
(k(q)\,a)-\Lambda_{c\nu}(q') \over \nu U/4t}\Bigr) \, ,
\label{Tapco1}
\end{eqnarray}

\begin{eqnarray}
k_{c\nu}(q) & = & q + {1\over \pi\,a}
\int_{q_c^{-}}^{q_c^{+}}\,dq'\,
N_c(q')\arctan\Bigl({\Lambda_{c\nu}(q)-\sin (k(q')\,a)\over \nu
U/4t}\Bigr) \nonumber \\ & + & {1\over 2\pi\,a}\sum_{\nu'
=1}^{\infty} \int_{-q_{c\nu '}}^{q_{c\nu'}}\,dq'\, N_{c\nu'}(q')
\Theta_{\nu,\,\nu'}\Bigl({\Lambda_{c\nu}(q)-\Lambda_{c\nu'}(q')
\over U/4t}\Bigr) \, , \label{Tapco2}
\end{eqnarray}
and

\begin{eqnarray}
0 & = & q - {1\over \pi\,a} \int_{q_c^{-}}^{q_c^{+}}\,dq'\,
N_c(q')\arctan\Bigl({\Lambda_{s\nu}(q)-\sin (k(q')\,a)
\over \nu U/4t}\Bigr)\nonumber \\
& + & {1\over 2\pi\,a}\sum_{\nu' =1}^{\infty} \int_{-q_{s\nu
'}}^{q_{s\nu'}}\,dq'\, N_{s\nu'}(q')\Theta_{\nu, \,\nu'}
\Bigl({\Lambda_{s\nu}(q)-\Lambda_{s\nu'}(q')\over U/4t}\Bigr) \, .
\label{Tapco3}
\end{eqnarray}
Here

\begin{equation}
k_{c\nu}(q) = {2\over a}\,{\rm Re}\,\Bigl\{\arcsin (\Lambda_{c\nu}
(q) + i \nu U/4t)\Bigr\} \, , \label{kcnco}
\end{equation}
is the $c\nu$ rapidity-momentum functional and the function
$\Theta_{\nu,\, \nu'} (x)$ is defined in Eq. (\ref{Theta}) of
Appendix B. Equations (\ref{Tapco1})-(\ref{Tapco3}) apply to all
regular energy eigenstates. The ground state and the low-energy
eigenstates involve occupancy configurations of the $c$ and $s1$
pseudoparticle branches only \cite{Carmelo92}. The pseudoparticles
have energy dispersions and residual interactions controlled by
$f$ functions, as the Fermi-liquid quasi-particles
\cite{Carmelo97,Carmelo92,II}. However, in contrast to such
quasi-particles, the pseudoparticles do not carry the same charge
and spin as the original electrons. The ground-state
pseudoparticle momentum distribution functions and the
pseudoparticle energy dispersions are presented in Appendix C.
Often the low-energy excitations generated by the occupancy
configurations of the $c$ and $s1$ pseudoparticle branches are
identified with holons and spinons, respectively
\cite{spectral0,Anderson,Penc}.

In this paper we find that the missing link between the original
electrons and the pseudoparticles whose occupancy configurations
describe the energy eigenstates are the rotated electrons. Rotated
electrons are related to the original electrons by a unitary
transformation. Such a transformation was considered for large
values of the on-site Coulombian repulsion $U$ in Ref.
\cite{Harris}. Interestingly, we find that the above $c\nu$ and
$s\nu$ pseudoparticles are composite $2\nu$-holon and
$2\nu$-spinon quantum objects, respectively. This leads to a
description where all energy eigenstates of the 1D Hubbard model
can be described in terms of occupancy configurations of three
elementary quantum objects: the spinon, the holon, and the $c$
pseudoparticle. We find that such elementary objects and the
rotated electrons are related. Within our general description the
energy eigenstates which are not associated with the Bethe-ansatz
solution, are described by holons and spinons that are not part of
the composite $c\nu$ and $s\nu$ pseudoparticles, respectively.
Interestingly, the latter holons and spinons are invariant under
the electron - rotated-electron unitary transformation.

As the quarks are confined within the composite nucleons
\cite{review}, the elementary spin $1/2$ spinons, $\eta$-spin
$1/2$ holons, and $c$ pseudoparticles are confined within the
present many-electron system. Indeed, only electrons can be
created to or annihilated from such a system. However, within the
non-perturbative many-electron system the electronic degrees of
freedom are organized in terms of the elementary spinons, holons,
and $c$ pseudoparticles. All energy eigenstates can be described
as occupancy configurations of these three elementary quantum
objects only.

In this paper we also address the important problem of the
transport of charge and spin. We find that the electronic degrees
of freedom couple to charge (and spin) probes through the holons
and $c$ pseudoparticles (and spinons). Those are the carriers of
charge (and spin) of the model for {\it all} energy scales. The
quantum-object description introduced in this paper is used in
Ref. \cite{V} in the construction of a theory for evaluation of
finite-energy few-electron spectral functions. That theory is
applied in Refs. \cite{spectral0,spectral} to the study of the
finite-energy one-electron spectral properties of low-dimensional
materials, as further discussed in Sec. V.

The paper is organized as follows: In Sec. II we describe the
non-perturbative organization of the electronic degrees of
freedom. The confirmation of the validity of the holon, spinon,
and $c$ pseudoparticle description is the subject of Sec. III. In
Sec. IV we study the problem of charge and spin transport.
Finally, the summary and concluding remarks are presented in Sec.
V.

\section{NON-PERTURBATIVE ORGANIZATION OF THE ELECTRONIC DEGREES OF FREEDOM}

Each lattice site $j=1,2,...,N_a$ of the model (\ref{H}) can
either be doubly occupied, empty, or singly occupied by a
spin-down or spin-up electron. The maximum number of electrons is
$2N_a$ and corresponds to density $n=2/a$. Besides the $N$
electrons, it is useful to consider $[2N_a-N]$ {\it electronic
holes}. (Here we use the designation {\it electronic hole} instead
of {\it hole}, in order to distinguish this type of hole from the
{\it pseudoparticle hole}, which appears later in this paper.) Our
definition of electronic hole is such that when a lattice site is
empty, we say that it is occupied by two electronic holes. If a
lattice site is singly occupied, we say that it is occupied by an
electron and an electronic hole. If a lattice site is doubly
occupied, it is unoccupied by electronic holes. Thus, in our
language an empty site corresponds to an occupancy of two
electronic holes at the same site $j$. We denote by $N^h$ the
total number of electronic holes,

\begin{equation}
N^h = [2N_a - N] \, . \label{Nh}
\end{equation}

One can describe charge and spin transport either in terms of
electrons or of electronic holes. In Sec. IV we discuss these two
alternative schemes. The studies of this paper use often the spin
$\sigma$ electron description. The eigenvalues of the diagonal
generators (\ref{Sz}) can be written in terms of the numbers
$N^h$, $N$, $N_{\uparrow}$, and $N_{\downarrow}$ as follows,

\begin{equation}
S_c^z = -{1\over 4}[N^h-N] \, ; \hspace{1cm} S_s^z = -{1\over
2}[N_{\uparrow}- N_{\downarrow}] \, . \label{Scsz}
\end{equation}

The pseudoparticle representation is extracted from the
Bethe-ansatz solution, yet that solution provides no explicit
information about the relation of these quantum objects to the
original electrons. In this section we introduce a holon and
spinon description which emerges from the electron -
rotated-electron unitary transformation. In the ensuing section,
we confirm the relation of rotated electrons to holons, spinons,
and $c$ pseudoparticles provided here.

\subsection{THE ELECTRON - ROTATED-ELECTRON UNITARY TRANSFORMATION}

We start by presenting a brief description of the concept of
rotated electron. The electron - rotated-electron unitary
transformation maps the electrons onto rotated electrons such that
rotated-electron double occupation, no occupation, and spin-up and
spin-down single occupation are good quantum numbers for {\it all}
values of $U/t$. Thus, there is at least a complete set of common
energy and rotated-electron double occupation, no occupation, and
spin-up and spin-down single occupation eigenstates. As mentioned
in Sec. I, until now a complete set of such states for all values
of $U/t$ was not found and this issue remains an interesting open
question.

Let us introduce the electron - rotated-electron unitary
transformation. We call $c_{j,\,\sigma}^{\dag}$ the electrons that
occur in the 1D Hubbard model (\ref{H}) and (\ref{HH}), while the
operator ${\tilde{c}}_{j,\,\sigma}^{\dag}$ such that,

\begin{equation}
{\tilde{c}}_{j,\,\sigma}^{\dag} =
{\hat{V}}^{\dag}(U/t)\,c_{j,\,\sigma}^{\dag}\,{\hat{V}}(U/t) \, ,
\label{c+til}
\end{equation}
represents the rotated electrons, where the electron -
rotated-electron unitary operator ${\hat{V}}(U/t)$ is defined
below. Inversion of the relation (\ref{c+til}) leads to
$c_{j,\,\sigma}^{\dag} =
{\hat{V}}(U/t)\,{\tilde{c}}_{j,\,\sigma}^{\dag}\,{\hat{V}}^{\dag}(U/t)$.
The rotated-electron double occupation operator is given by,

\begin{equation}
\tilde{D} \equiv {\hat{V}}^{\dag}(U/t)\,\hat{D}\,{\hat{V}}(U/t) =
\sum_{j}\, {\tilde{c}}_{j,\,\uparrow }^{\dagger
}\,{\tilde{c}}_{j,\,\uparrow }\, {\tilde{c}}_{j,\,\downarrow
}^{\dagger }\,{\tilde{c}}_{j,\,\downarrow } \, , \label{Doptil}
\end{equation}
where $\hat{D}$ is the electron double occupation operator given
in Eq. (\ref{Dop}). Note that $c_{j,\,\sigma}^{\dag}$ and
${\tilde{c}}_{j,\,\sigma}^{\dag}$ are only identical in the
$U/t\rightarrow\infty$ limit where electron double occupation
becomes a good quantum number.

The expression of any rotated operator $\tilde{O} =
{\hat{V}}^{\dag}(U/t)\,\hat{O}\,{\hat{V }}(U/t)$ in terms of the
rotated electron operators ${\tilde{c}}_{j,\,\sigma}^{\dag}$  and
${\tilde{c}}_{j,\,\sigma}$ is the same as the expression for the
corresponding general operator $\hat{O} =
{\hat{V}}(U/t)\,\tilde{O}\,{\hat{V}}^{\dag}(U/t)$ in terms of the
electron operators $c_{j,\,\sigma}^{\dag}$ and $c_{j,\,\sigma}$,
respectively. The operators ${\hat{V}}^{\dag}(U/t)$ and
${\hat{V}}(U/t)$ associated with the electron - rotated-electron
unitary transformation can be written as,

\begin{equation}
{\hat{V}}^{\dag}(U/t) = e^{-\hat{S}} \, ; \hspace{0.5cm}
{\hat{V}}(U/t) = e^{\hat{S}} \, . \label{SV}
\end{equation}
The operator $\hat{S}$ of Eq. (\ref{SV}) is uniquely defined by
the following two equations,

\begin{equation}
{\tilde{H}}_H = {\hat{V}}^{\dag}(U/t)\,{\hat{H}}_H\,{\hat{V}}(U/t)
= {\hat{H}}_H + [{\hat{H}}_H,\,{\hat{S}}\,] + {1\over
2}\,[[{\hat{H}}_H,\,{\hat{S}}\,],\,{\hat{S}}\,] + ... \,
,\label{HHtil}
\end{equation}
and

\begin{equation}
[{\hat{H}}_H,\,{\hat{V}}^{\dag}(U/t)\,\hat{D}\,{\hat{V}}(U/t)] =
[{\hat{H}}_H,\,\tilde{D}] = 0 \, , \label{HHDtil}
\end{equation}
where the Hamiltonian ${\hat{H}}_H$ and the rotated-electron
double occupation operator $\tilde{D}$ are given in Eqs.
(\ref{HH}) and (\ref{Doptil}), respectively.

The transformation associated with the electron - rotated-electron
unitary operator ${\hat{V}}(U/t)$ was introduced in Ref.
\cite{Harris}. The studies of that reference referred to large
values of $U/t$ and did not clarify for arbitrary values of $U/t$
the relation of rotated-electron double occupation to the quantum
numbers provided by the Bethe-ansatz solution. However, this
transformation is uniquely defined for all values of $U/t$ by Eqs.
(\ref{SV})-(\ref{HHDtil}). Equations (\ref{HHtil}) and
(\ref{HHDtil}) can be used to derive an expression for the unitary
operator order by order in $t/U$. The authors of the second paper
of Ref. \cite{Harris} carried out this expansion up to eighth
order (see foot note [12]).

\subsection{THE HOLON, SPINON, AND $c$ PSEUDOPARTICLE
FLUIDS}

One should distinguish the total $\eta$ spin (and spin) value,
which we denote by $S_c$ (and $S_s$) and the corresponding
$\eta$-spin (and spin) projection, which we denote by $S_c^z$ (and
$S_s^z$) from the $\eta$ spin (and spin) carried by the elementary
quantum objects. We call $s_c$ (and $s_s$) the $\eta$ spin (and
spin) carried by the holons, spinons, and other elementary objects
and $\sigma_c$ (and $\sigma_s$) their $\eta$-spin (and spin)
projection. The operators ${\hat{M}}_{c,\,\sigma_{c}}$ and
${\hat{M}}_{s,\,\sigma_{s}}$, which count the number of the
$\sigma_{c}=\pm 1/2$ holons and $\sigma_{s}=\pm 1/2$ spinons, have
the following form,

\begin{eqnarray}
{\hat{M}}_{c,\,-1/2} & = & {\tilde{R}}_{c,\,-1} =
{\hat{V}}^{\dag}(U/t)\,\sum_{j}
c_{j\uparrow}^{\dag}\,c_{j\uparrow}\,
c_{j\downarrow}^{\dag}\,c_{j\downarrow}\,{\hat{V}}(U/t) \, ; \nonumber \\
{\hat{M}}_{c,\,+1/2} & = & {\tilde{R}}_{c,\,+1} =
{\hat{V}}^{\dag}(U/t)\,\sum_{j}
c_{j\uparrow}\,c_{j\uparrow}^{\dag}\,
c_{j\downarrow}\,c_{j\downarrow}^{\dag}\,{\hat{V}}(U/t)
\, ; \nonumber \\
{\hat{M}}_{s,\,-1/2} & = & {\tilde{R}}_{s,\,-1} =
{\hat{V}}^{\dag}(U/t)\,\sum_{j}
c_{j\downarrow}^{\dag}\,c_{j\uparrow}\,
c_{j\uparrow}^{\dag}\,c_{j\downarrow}\,{\hat{V}}(U/t) \, ; \nonumber \\
{\hat{M}}_{s,\,+1/2} & = & {\tilde{R}}_{s,\,+1}  =
{\hat{V}}^{\dag}(U/t)\,\sum_{j}
c_{j\uparrow}^{\dag}\,c_{j\downarrow}\,
c_{j\downarrow}^{\dag}\,c_{j\uparrow}\,{\hat{V}}(U/t) \, .
\label{Mcs-+}
\end{eqnarray}
Here ${\hat{V}}(U/t)$ is the electron - rotated-electron unitary
operator uniquely defined for all values of $U/t$ by Eqs.
(\ref{SV})-(\ref{HHDtil}) and the operators,

\begin{eqnarray}
{\hat{R}}_{c,\,-1} & \equiv & \hat{D} = \sum_{j}
c_{j,\,\uparrow}^{\dag}\,c_{j,\,\uparrow}
c_{j,\,\downarrow}^{\dag}\,c_{j,\,\downarrow} \, ; \hspace{1cm}
{\hat{R}}_{c,\,+1} = \sum_{j}
c_{j,\,\uparrow}\,c_{j,\,\uparrow}^{\dag}\,
c_{j,\,\downarrow}\,c_{j,\,\downarrow}^{\dag} \, , \label{Rc}
\end{eqnarray}
and

\begin{eqnarray}
{\hat{R}}_{s,\,-1} & = & \sum_{j}
c_{j,\,\downarrow}^{\dag}\,c_{j\uparrow}
c_{j,\,\uparrow}^{\dag}\,c_{j,\,\downarrow} \, ; \hspace{1cm}
{\hat{R}}_{s,\,+1} = \sum_{j}
c_{j,\,\uparrow}^{\dag}\,c_{j,\,\downarrow}\,
c_{j,\,\downarrow}^{\dag}\,c_{j,\,\uparrow} \, , \label{Rs}
\end{eqnarray}
are such that ${\hat{R}}_{c,\,-1}$ counts the number of electron
doubly-occupied sites, ${\hat{R}}_{c,\,+1}$ counts the number of
electron empty sites, ${\hat{R}}_{s,\,-1}$ counts the number of
electron down-spin singly-occupied sites, and ${\hat{R}}_{s,\,+1}$
counts the number of electron up-spin singly-occupied sites.

The new physics brought about by the relations of Eq.
(\ref{Mcs-+}) is that the Hilbert-space rotation performed by the
unitary operator ${\hat{V}}(U/t)$ produces new exotic quantum
objects whose occupancy configurations describe the exact energy
eigenstates. The unitary rotation is such that for all values of
$U/t$ the number of emerging $-1/2$ holons, $+1/2$ holons, $-1/2$
spinons, and $+1/2$ spinons equals precisely the number of
rotated-electron doubly occupied sites, empty sites, spin-down
singly occupied sites, and spin-up singly occupied sites,
respectively. In Sec. III we confirm that the $\pm 1/2$ holon and
$\pm 1/2$ spinon numbers introduced in Eq. (\ref{Mcs-+}) are
consistent with the numbers of $c$ pseudoparticles and $\alpha\nu$
pseudoparticles where $\alpha =c,\,s$ and $\nu=1,2,...$ obtained
from the Bethe-ansatz solution.

The first step of the organization of the electronic degrees of
freedom which arises from the non-perturbative effects of the
many-electron interactions is the Hilbert-space rotation that maps
electrons onto rotated electrons. This involves a decoupling of
the $N$-electron system in a $N_c$-rotated-electron and a
$[N-N_c]$-rotated-electron fluids. (The number of electrons equals
the number of rotated electrons.) On the one hand, $N_c$ is
nothing but the number of $c$ pseudoparticles introduced in
Appendix B from Takahashi's thermodynamic Bethe-ansatz equations.
On the other hand, we identify the number $N_c$ with the number of
rotated-electron singly occupied sites. It follows that the number
$[N-N_c]/2$ equals the number of rotated-electron doubly occupied
sites. This is consistent with the fact that $[N-N_c]$ is always
an even number. Moreover, the $N^h$-electronic hole system
decouples into a $N_c$-rotated-electronic hole and a
$[N^h-N_c]$-rotated-electronic hole fluids, where the number
$[N^h-N_c]$ is also even. The number $N_c$ also equals the number
of rotated-electronic holes associated with the $N_c$
rotated-electron singly occupied sites. Furthermore, the above
$[N^h-N_c]$-rotated-electronic hole fluid corresponds to the
$[N^h-N_c]/2$ rotated-electron empty sites. The
$N_c$-rotated-electron (and $N_c$-rotated-electronic hole) fluid
corresponds to $N_c$ {\it unpaired} rotated electrons (and rotated
electronic holes), whereas the $[N-N_c]$-rotated-electron fluid
(and $[N^h-N_c]$-rotated-electronic hole fluid) is described by
$[N-N_c]/2$ (and $[N^h-N_c]/2$) quantum objects described by
on-site pairs of rotated electrons (and on-site pairs of
rotated-electronic holes).

The $[N-N_c]$-rotated-electron fluid involves $[N-N_c]/2$
spin-down rotated electrons and an equal number of spin-up rotated
electrons which {\it pair} and form $[N-N_c]/2$ spin zero singlet
rotated-electron on-site pairs associated with the doubly occupied
sites. Each of these rotated-electron on-site pairs has $s_c=1/2$,
$s_s=0$, and $\sigma_c= -1/2$ and is identified with a
$\sigma_c=-1/2$ holon or, simply, a $-1/2$ holon. Thus, the $-1/2$
holons correspond to the rotated-electron doubly occupied sites,
consistently to Eqs. (\ref{Mcs-+})-(\ref{Rs}). Moreover, the two
rotated-electronic holes of the $[N^h-N_c]$-rotated-electronic
hole fluid are also {\it paired} in the same empty side. Each of
these $[N^h-N_c]/2$ rotated-electronic hole pairs has $s_c=1/2$,
$s_s=0$, and $\sigma_c = +1/2$ and is called a $\sigma_c=+1/2$
holon or, simply, a $+1/2$ holon. Such $\sigma_c=+1/2$ holons
correspond to the rotated-electron empty sites, again consistently
with the relations given in Eqs. (\ref{Mcs-+})-(\ref{Rs}). We
emphasize that the total number $[N-N_c]/2+[N^h-N_c]/2=[N_a-N_c]$
of holons equals the number $N^h_c=[N_a-N_c]$ of $c$
pseudoparticle holes, as discussed below.

The $N_c$-rotated-electron fluid is characterized by an exotic
separation of the spin and charge degrees of freedom. Such a
decoupling leads to a $N_c$ spinon fluid and a $N_c$ $c$
pseudoparticle fluid. The $c$ pseudoparticle has no spin degrees
of freedom. It corresponds to the charge part of a
rotated-electron singly occupied site. There is a {\it local $c$
pseudoparticle} description \cite{V} associated with the momentum
$c$ pseudoparticle representation obtained from the Bethe-ansatz
solution in Appendix B. It is the local $c$ pseudoparticle which
corresponds to the charge part of a rotated-electron singly
occupied site. The local and momentum pseudoparticle descriptions
are related by a simple Fourier transform \cite{V}. We call the
charge part of the rotated electron of such a site a {\it
chargeon}. In case of charge transport in terms of electronic
holes, one should introduce the {\it antichargeon}. This is the
charge part of the rotated-electronic hole of a singly occupied
site. Furthermore, the spin-projection $\pm 1/2$ spinon
corresponds to the spin part of a spin-projection $\pm 1/2$
rotated-electron singly occupied site. The $c$ pseudoparticle
excitations describe the charge motion of the rotated-electron
singly occupied sites relative to the rotated-electron
doubly-occupied and empty sites. Moreover, the
$N_c$-rotated-electronic hole fluid corresponds to the $N_c$
antichargeons. Thus, the $c$ pseudoparticle includes a chargeon
and an antichargeon. However, we emphasize that in what charge
transport is concerned, the chargeon and antichargeon correspond
to alternative descriptions of the $c$ pseudoparticle, as
discussed in Sec. IV. Also spin transport can either be described
in terms of electrons or electronic holes, as discussed in the
same section. In general we consider in this paper the description
of charge and spin transport in terms of electrons. For electron
spin transport, the number of spinons with spin projections
$\sigma_s=+1/2$ and $\sigma_s=-1/2$ equals the number of spin-up
rotated electrons $[N_c + N_{\uparrow}-N_{\downarrow}]/2$ and the
number of spin-down rotated electrons $[N_c +
N_{\downarrow}-N_{\uparrow}]/2$ in the $N_c$-rotated-electron
fluid, respectively. The numbers $[N_c +
N_{\uparrow}-N_{\downarrow}]/2$ and $[N_c +
N_{\downarrow}-N_{\uparrow}]/2$ also equal the numbers of singly
occupied sites occupied by spin-up and spin-down rotated
electrons, respectively.

Let us summarize the above expressions for the numbers of the
elementary quantum objects in a few equations. Consistently with
the notation used in the relations given in Eq. (\ref{Mcs-+}), we
call $M_{c,\,\sigma_c}$ the number of $\sigma_c$ holons and the
total number of holons is denoted by $M_c
=\sum_{\sigma_c}\,M_{c,\,\sigma_c}$. These values read,

\begin{equation}
M_{c,\,-1/2} = {1\over 2}\Bigl[N - N_c\Bigr] \, ; \hspace{0.5cm}
M_{c,\,+1/2} = {1\over 2}\Bigl[N^h - N_c\Bigr] \, ; \hspace{0.5cm}
M_c = {1\over 2}\Bigl[N^h + N\Bigr] -N_c = [N_a -N_c] \, .
\label{Mcmp}
\end{equation}

On the other hand, we call $M_{s,\,\sigma_s}$ the number of
$\sigma_s$ spinons and $M_s =\sum_{\sigma_s}\,M_{s,\,\sigma_s}$
the total number of spinons. These numbers are given by

\begin{equation}
M_{s,\,-1/2} = {1\over 2}\Bigl[N_c - N_{\uparrow} +
N_{\downarrow}\Bigr] \, ; \hspace{0.5cm} M_{s,\,-1/2} = {1\over
2}\Bigl[N_c + N_{\uparrow} - N_{\downarrow}\Bigr] \, ;
\hspace{0.5cm} M_s=N_c \, . \label{Nud}
\end{equation}
The numbers $M_{c,\,\sigma_c}$, $M_{s,\,\sigma_s}$, and $N_c$ of
the elementary quantum objects are good quantum numbers.

The following relations between the numbers of $\pm 1/2$ holons,
$\pm 1/2$ spinons, and $c$ pseudoparticles are valid for the whole
Hilbert space of the model,

\begin{equation}
M_c = \sum_{\sigma_c=\pm 1/2}\, M_{c,\,\sigma_c} = N_c^h =
[N_a-N_c] \, ; \hspace{1cm} M_s = \sum_{\sigma_s=\pm 1/2}\,
M_{s,\,\sigma_s} = N_c \, ; \hspace{1cm} \sum_{\alpha
=c,s}\,M_{\alpha} = N_a \, . \label{Mcs}
\end{equation}

The total number of holons plus the total number of spinons equals
the number of lattice sites $N_a$. This result follows from the
one-to-one correspondence between the holon and spinon numbers and
the numbers of rotated electrons. Since $N_a$ is even, such a sum
rule implies that the symmetry of the 1D Hubbard model is $SO(4)$,
consistently with the results of Refs. \cite{HL,Yang89}. Indeed,
half of the irreducible representations of $SU(2)\times SU(2)$ are
excluded by the constraint imposed by such a sum rule: States
where the total number of holons plus spinons is odd are excluded.

\subsection{THE $2\nu$-HOLON AND $2\nu$-SPINON COMPOSITE PSEUDOPARTICLES}

The $c\nu$ pseudoparticles (and $s\nu$ pseudoparticles) obtained
from the Bethe-ansatz solution in Appendix B and whose energy
bands are given in Appendix C are $2\nu$-holon (and $2\nu$-spinon)
composite objects. Such composite pseudoparticles involve an equal
number $\nu=1,2,...$ of $+1/2$ holons and $-1/2$ holons (and
$+1/2$ spinons and $-1/2$ spinons). The composite $c\nu$
pseudoparticle has $s_c=0$ and $s_s=0$ and the composite $s\nu$
pseudoparticle has $s_s=0$ and no charge degrees of freedom.

As for the case of the $c$ pseudoparticles, there is a {\it local
$\alpha\nu$ pseudoparticle} description associated with the
momentum $\alpha\nu$ pseudoparticle representation obtained from
the Bethe-ansatz solution in Appendix B \cite{V}. The local $c\nu$
pseudoparticle (and $s\nu$ pseudoparticle) is a $2\nu$-holon (and
$2\nu$-spinon) composite quantum object of $\nu$ $-1/2$ holons and
$\nu$ $+1/2$ holons (and $\nu$ $-1/2$ spinons and $\nu$ $+1/2$
spinons) associated with $\nu$ rotated-electron doubly-occupied
and empty sites (and $\nu$ spin-down and spin-up rotated-electron
singly occupied sites), respectively. The local and momentum
$\alpha\nu$ pseudoparticle descriptions are related by a simple
Fourier transform \cite{V}. The momentum $c\nu$ pseudoparticle
(and $s\nu$ pseudoparticle) excitations are associated with hoping
rotated-electron processes within the doubly-occupied and empty
site (and singly-occupied site) subsystem. The momentum of these
excitations corresponds to the third term (and second term) of Eq.
(\ref{Ppco}). On the other hand, the momentum $c$ pseudoparticle
excitations correspond to relative movements of the rotated
electrons associated with these two subsystems. These excitations
conserve both the rotated-electron internal doubly-occupied and
empty site occupancy configurations and the singly-occupied site
occupancy configurations. The momentum of these excitations
corresponds to the first term of Eq. (\ref{Ppco}).

The numbers of $\pm 1/2$ holons ($\alpha =c$) and of $\pm 1/2$
spinons ($\alpha =s$) can be written as,

\begin{equation}
M_{\alpha,\,\pm 1/2} = L_{\alpha ,\,\pm 1/2} + \sum_{\nu
=1}^{\infty}\nu\,N_{\alpha\nu} \, ; \hspace{1cm} \alpha = c,s \, ,
\label{Mas}
\end{equation}
where the term $\sum_{\nu =1}^{\infty}\nu\,N_{\alpha\nu}$ on the
right-hand side of this equation gives the number of $\pm 1/2$
holons ($\alpha =c$) or of $\pm 1/2$ spinons ($\alpha =s$) that
are part of the $2\nu$-holon composite $c\nu$ pseudoparticles or
$2\nu$-spinon composite $s\nu$ pseudoparticles, respectively. On
the right-hand side of Eq. (\ref{Mas}) $N_{\alpha\nu}$ gives the
corresponding number of $\alpha\nu$ pseudoparticles and

\begin{equation}
L_{\alpha ,\,{\sigma}_{\alpha}} = S_{\alpha} -2{\sigma}_{\alpha}\,
S_{\alpha}^z = L_{\alpha}/2 -2{\sigma}_{\alpha}\, S_{\alpha}^z \,
; \hspace{1cm} \alpha = c,s \, , \label{Las}
\end{equation}
is the number of $\sigma_c=\pm 1/2$ holons ($\alpha =c$) or of
$\sigma_s=\pm 1/2$ spinons ($\alpha =s$) which are not part of
composite $c\nu$ or $s\nu$ pseudoparticles, respectively.

Interestingly, the spin singlet and composite two-spinon character
of the $s1$ pseudoparticles reveals that the quantum spin fluid
described by these quantum objects is the 1D realization of the
two-dimensional {\it resonating valence bond} (RVB) spin fluid
\cite{Anderson}. In Sec. IV we find that the charge (and spin)
quantum fluid described by the $c$ pseudoparticles and $c\nu$
pseudoparticles (and $s\nu$ pseudoparticles) plays a central role
in the charge (and spin) transport properties. Indeed, the holons
(and spinons) which are not part of $c\nu$ pseudoparticles (and
$s\nu$ pseudoparticles) are localized quantum objects and do not
contribute to charge (and spin) transport \cite{II}. In the limit
$U/t\rightarrow\infty$ both the $c\nu$ pseudoparticles and $s\nu$
pseudoparticles also become localized quantum objects and only the
$c$ pseudoparticles contribute to transport of charge \cite{II}.
In this limit the $c$ pseudoparticles are related to the spin-less
fermions of Ref. \cite{Penc}.

\subsection{THE YANG HOLONS, HL SPINONS, AND THE CORRESPONDING TWO
$SU(2)$ ALGEBRAS}

We call $\pm 1/2$ {\it Yang holons} (and $\pm 1/2$ {\it HL
spinons}) the $\pm 1/2$ holons (and $\pm 1/2$ spinons) which are
not part of composite $c\nu$ pseudoparticles (and $s\nu$
pseudoparticles) and whose numbers are given by Eq. (\ref{Las})
with $\alpha =c$ (and $\alpha =s$). In the case of the HL spinons,
HL stands for Heilmann and Lieb who are the authors of Ref.
\cite{HL}, whereas in the case of the Yang holons, Yang stands for
C. N. Yang who is the author of Ref. \cite{Yang89}. In these
references it was found that in addition to the $SU(2)$ spin
symmetry, the Hubbard model has a $SU(2)$ $\eta$-spin symmetry.

An important physical property of the $\pm 1/2$ Yang holons and
$\pm 1/2$ HL spinons is that these quantum objects remain
invariant under the electron - rotated-electron unitary
transformation for all values of $U/t$. This is confirmed by the
studies of Ref. \cite{II}, where it is found that $\pm 1/2$ Yang
holons (and $\pm 1/2$ HL spinons) refer to electron
doubly-occupied sites (and spin $\pm 1/2$ electron singly occupied
sites). Although  for finite values of $U/t$ electron double
occupation is not a good quantum number, these particular quantum
objects have a local character such that these concepts apply to
their energy-eigenstate site occupancy configurations. It follows
that the Yang holon (and HL spinon) and the rotated Yang holon
(and rotated HL spinon) are the same quantum object for all values
of $U/t$. This is in contrast to the pseudoparticles which become
invariant under the electron - rotated-electron unitary
transformation as $U/t\rightarrow\infty$ only \cite{II}.
Otherwise, the pseudoparticle and the rotated pseudoparticle are
in general different quantum objects.

The Yang holons (and HL spinons) also have a very different
behavior relative to holons (and spinons) which are part of
composite $c\nu$ pseudoparticles (and $s\nu$ pseudoparticles) in
what the $\eta$-spin (and spin) $SU(2)$ algebra is concerned.
Indeed, the designations of Yang holons and HL spinons is
justified by the fact that out of a total number $M_{\alpha,\,\pm
1/2}$ of $\pm 1/2$ holons ($\alpha =c$) or $\pm 1/2$ spinons
($\alpha =s$), only the $L_{c,\,\pm 1/2}$ $\pm 1/2$ holons or
$L_{s,\,\pm 1/2}$ $\pm 1/2$ spinons which are not part of
$\alpha\nu$ pseudoparticles, are sensitive to the application of
the $\eta$-spin ($\alpha =c$) and spin ($\alpha =s$) off-diagonal
generators, respectively. This results from the fact that the
composite $\alpha\nu$ pseudoparticles are $\eta$-spin ($\alpha=c$)
and spin ($\alpha=s$) singlet combinations of $2\nu$ holons
($\alpha=c$) and $2\nu$ spinons ($\alpha=s$) and thus have
$s_{\alpha}=0$, {\it i.e.} have zero $\eta$ spin ($\alpha= c$) or
zero spin ($\alpha =s$).

The off-diagonal generators of the $\eta$-spin algebra, Eq.
(\ref{Sc}), transform $-1/2$ Yang holons into $+1/2$ Yang holons
and vice versa. Also the off-diagonal generators of the spin
algebra, Eq. (\ref{Ss}), transform $-1/2$ HL spinons into $+1/2$
HL spinons and vice versa.  Thus, these generators produce
$\eta$-spin and spin flips, respectively. Application of
${\hat{S}}_c^{\dagger}$ (or ${\hat{S}}_s^{\dagger}$) produces a
$\eta$-spin flip (or spin flip) which transforms a $+1/2$ Yang
holon (or $+1/2$ HL spinon) into a $-1/2$ Yang holon (or $-1/2$ HL
spinon). Application of ${\hat{S}}_c$ (or ${\hat{S}}_s$) produces
a $\eta$-spin flip (or spin flip) which transforms a $-1/2$ Yang
holon (or $-1/2$ HL spinon) into a $+1/2$ Yang holon (or $+1/2$ HL
spinon). The LWSs of the $\eta$-spin (or spin) algebra, have zero
occupancies of $-1/2$ Yang holons (or of $-1/2$ HL spinons).
Similarly, the HWSs of the $\eta$-spin (or spin) algebra have zero
occupancies of $+1/2$ Yang holons (of $+1/2$ HL spinons). Our
description corresponds to the whole Hilbert space of the 1D
Hubbard model, and includes states with both $-1/2$ and $+1/2$
Yang holon finite occupancies and with both $-1/2$ and $+1/2$ HL
spinon finite occupancies.

The electron - rotated-electron unitary operator ${\hat{V}}(U/t)$
commutes with all three generators of the $\eta$-spin (and spin)
algebra. It is this symmetry that is behind the invariance under
the same transformation of the corresponding holons (and spinons)
which are sensitive to these generators. However, there is no such
a symmetry requirement for the holons (and spinons) whose
arrangements are insensitive to the application of the same
generators. This analysis is consistent with the transformation
laws of the associated composite $c\nu$ pseudoparticles (and
$s\nu$ pseudoparticles) which are not invariant under the electron
- rotated-electron unitary transformation. Such an interplay of
the electron - rotated-electron unitary transformation and
$\eta$-spin and spin $SU(2)$ symmetries is important for the
spectral properties \cite{V}. Creation and annihilation elementary
operators can be introduced for the $\alpha\nu$ pseudoparticles
\cite{V}. Consistently with the above separation, these operators
commute with the off-diagonal generators of the $\eta$-spin and
spin algebras.

From equations (\ref{Mas})-(\ref{Las}), the total number of holons
($\alpha =c$) and spinons ($\alpha =s$) given in Eq. (\ref{Mcs})
can be expressed as,

\begin{equation}
M_{\alpha} = 2S_{\alpha} + \sum_{\nu
=1}^{\infty}2\nu\,N_{\alpha\nu} = L_{\alpha} + \sum_{\nu
=1}^{\infty}2\nu\,N_{\alpha\nu} \, ; \hspace{1cm} \alpha = c,s  \,
, \label{Ma}
\end{equation}
where

\begin{equation}
L_{\alpha} = \sum_{\sigma_{\alpha}=\pm 1/2}\, L_{\alpha
,\,\sigma_{\alpha}} = 2S_{\alpha} \, ; \hspace{1cm} \alpha = c,s
\, , \label{La}
\end{equation}
is for $\alpha =c$ (and $\alpha =s$) the total number of Yang
holons (and HL spinons). Note that Eq. (\ref{La}) is consistent
with the application of the off-diagonal generators of the
$\eta$-spin and spin algebras producing Yang holon $\eta$-spin
flips and HL spinon flips, respectively. Let us consider a regular
energy eigenstate. Since in such a state there are $2S_c$ $+1/2$
Yang holons and $2S_s$ $+1/2$ HL spinons, one can apply the
generator ${\hat{S}}_c^{\dagger}$ $2S_c$ times and the generator
${\hat{S}}_s^{\dagger}$ $2S_s$ times. This leads to $\eta$-spin
and spin towers with $2S_c +1$ and $2S_s +1$ states, respectively.
Therefore, from each LWS of the $S_{\alpha}$ algebra, one can
generate $2S_{\alpha}$ non LWSs of that algebra, which is the
correct result.

We emphasize that the property that the generators of the
$\eta$-spin and spin $SU(2)$ algebra commute with the $c$
pseudoparticle and composite $\alpha\nu$ pseudoparticle creation
and annihilation operators has the following important effect: All
$2S_{\alpha}$ energy eigenstates obtained from a given regular
energy eigenstate have the same pseudoparticle momentum
distribution functions $N_c (q)$ and $N_{\alpha\nu}(q)$ for all
branches $\alpha =c,s$ and $\nu =1,2,...$. Thus, these states are
described by similar pseudoparticle occupancy configurations and
only differ in the relative numbers of $+1/2$ Yang holons and
$-1/2$ Yang holons ($\alpha =c$) or/and in the relative numbers of
$+1/2$ HL spinons and $-1/2$ HL spinons ($\alpha =s$). Thus, the
set of $2S_{\alpha}$ non-regular energy eigenstates belonging to
the same tower of states differ only in the values of the numbers
$L_{\alpha ,\,-1/2}$ and $L_{\alpha ,\,+1/2}$ such that $L_{\alpha
,\,-1/2}+L_{\alpha ,\,+1/2}=2S_{\alpha}$, yet the total number of
holons and spinons is the same for all of them. Importantly, this
confirms that the coupled functional equations
(\ref{Tapco1})-(\ref{Tapco3}), which involve the pseudoparticle
momentum distribution functions and do not depend on the
$L_{\alpha ,\,\sigma_{\alpha}}$ numbers, describe both regular and
non-regular energy eigenstates. The values of the
rapidity-momentum functional $k(q)$ and rapidity functionals
$\Lambda_{\alpha\nu}(q)$ are the same for all $2S_{\alpha}+1$
states in the same tower. Such functionals are eigenvalues of
operators which commute with the off-diagonal generators of
$\eta$-spin and spin algebras. This is consistent with the
$\eta$-spin and spin $SU(2)$ symmetries, which imply that the
Hamiltonian commutes with these generators and thus the energy
(\ref{Epso4co}) is the same for the set of $2S_{\alpha}+1$ states
belonging to the same $\eta$-spin ($\alpha =c$) or spin ($\alpha
=s$) tower. We recall that such an energy expression is fully
determined by the values of the rapidity-momentum and rapidity
functionals. These results reveal that the holon and spinon
description introduced in this paper permits the extension of the
functional equations (\ref{Tapco1})-(\ref{Tapco3}) to the whole
Hilbert space of the 1D Hubbard model.

From the use of Eqs. (\ref{Mas}) and (\ref{Las}) we find that the
$\eta$-spin value $S_c$, spin value $S_s$, and the corresponding
projections $S_c^z$ and $S_s^z$, respectively, can be expressed in
terms of the numbers of $\pm 1/2$ Yang holons and of $\pm 1/2$ HL
spinons as follows,

\begin{equation}
S_{\alpha} = {1\over 2}[L_{\alpha,\,+1/2}+ L_{\alpha,\,-1/2}] =
{1\over 2}M_{\alpha} - \sum_{\nu =1}^{\infty}\nu\,N_{\alpha\nu} \,
; \hspace{1cm} \alpha = c,s \, , \label{SM}
\end{equation}

\begin{eqnarray}
S_c^z & = & -{1\over 2}[L_{c,\,+1/2}- L_{c,\,-1/2}] = -{1\over
2}[M_{c,\,+1/2}- M_{c,\,-1/2}] \, ; \nonumber \\
S_s^z & = & -{1\over 2}[L_{s,\,+1/2}- L_{s,\,-1/2}] = -{1\over
2}[M_{s,\,+1/2}- M_{s,\,-1/2}] \, . \label{Szhs}
\end{eqnarray}
On the right-hand side of Eqs. (\ref{SM}) and (\ref{Szhs}) $S_c$,
$S_s$, $S_c^z$, and $S_s^z$ were also expressed in terms of the
numbers of holons, spinons, and pseudoparticles.

The $\eta$-spin and spin value expressions (\ref{Scco}),
(\ref{Ssco}), and (\ref{SM}) and the energy expression
(\ref{Epso4co}) are valid for the whole Hilbert space. In
contrast, we note that the momentum expression (\ref{Ppco}) refers
only to LWSs of the $\eta$-spin and spin $SU(2)$ algebras.
Generalization of that momentum expression to the whole Hilbert
space leads to,

\begin{equation}
P = {L\over 2\pi}\int_{q_c^{-}}^{q_c^{+}}\,dq\, N_c (q)\, q +
{L\over 2\pi}\sum_{\nu =1}^{\infty} \int_{-q_{s\nu
}}^{q_{s\nu}}\,dq\, N_{s\nu} (q)\, q + {L\over 2\pi}\sum_{\nu
=1}^{\infty}\int_{-q_{c\nu }}^{q_{c\nu }}\,dq\, N_{c\nu} (q)\,
[{\pi\over a} -q] + {\pi\over a}\,M_{c,\,-1/2} \, . \label{Ppco2}
\end{equation}
Here the contributions from the $c$ pseudoparticles, spinons (in
terms of composite $s\nu$ pseudoparticles), and holons lead to
different terms. Note that the term $[\pi/a]\,M_{c,\,-1/2}$ is
consistent with creation of a $-1/2$ Yang holon requiring momentum
$\pi/a$. The value of the $-1/2$ Yang holon momentum follows from
the form of the $\eta$-spin off-diagonal generators defined in Eq.
(\ref{Sc}). There is an additional contribution of momentum
$[\pi/a]\,\nu$ to the term $[\pi/a]\,M_{c,\,-1/2}$ and of momentum
$[\pi/a -q]$ to the third term of Eq. (\ref{Ppco2}) for each
$2\nu$-holon composite $c\nu$ pseudoparticle.

\section{CONFIRMATION OF THE HOLON, SPINON, AND $c$ PSEUDOPARTICLE DESCRIPTION}

The goals of this section are to confirm: (1) the validity of the
holon and spinon number operator expressions given in Eq.
(\ref{Mcs-+}); (2) that the $\alpha\nu$ pseudoparticle
representation provided by the Bethe-ansatz solution leads to the
same complete description of the LWSs as the $\pm 1/2$ holons and
$\pm 1/2$ spinons that are not invariant under the electron -
rotated-electron unitary transformation.

The concept of CPHS ensemble subspace where CPHS stands for $c$
pseudoparticle, holon, and spinon \cite{II} is used in this
section. This is a Hilbert subspace spanned by all states with
fixed values for the $-1/2$ Yang holon number $L_{c,\,-1/2}$,
$-1/2$ HL spinon number $L_{s,\,-1/2}$, $c$ pseudoparticle number
$N_c$, and for the sets of $\alpha\nu$ pseudoparticle numbers
$\{N_{c\nu}\}$ and $\{N_{s\nu}\}$ corresponding to the
$\nu=1,2,...$ branches.

\subsection{RELATION OF THE HOLONS, SPINONS, AND $c$ PSEUDOPARTICLES
TO ROTATED ELECTRONS}

Let us consider the four expectation values
$R_{\alpha,\,l_{\alpha}}=\langle
{\hat{R}}_{\alpha,\,l_{\alpha}}\rangle$, where $\alpha =c,\,s$ and
$l_{\alpha}=-1,\,+1$. The corresponding operator
${\hat{R}}_{c,\,-1}$ counts the number of electron doubly-occupied
sites, ${\hat{R}}_{c,\,+1}$ counts the number of electron empty
sites, ${\hat{R}}_{s,\,-1}$ counts the number of spin-down
electron singly-occupied sites, and ${\hat{R}}_{s,\,+1}$ counts
the number of spin-up electron singly-occupied sites. These
operators are given in Eqs. (\ref{Rc}) and (\ref{Rs}) and obey the
relations (56) and (57) of Ref. \cite{II}. These operational
relations are valid for the whole parameter space and reveal that
in a given electronic ensemble space, out of the four expectation
values of the operators of Eqs. (\ref{Rc}) and (\ref{Rs}), only
one is independent. The electron double-occupation expectation
value $D\equiv R_{c,\,-1}$ was studied in Ref. \cite{II}. The use
of the operational Eqs. (56) and (57) of the same reference
provides the corresponding values for $R_{c,\,+1}$, $R_{s,\,-1}$,
and $R_{s,\,+1}$.

To start with we consider that the $\sigma_c=\pm 1/2$ holon and
$\sigma_s=\pm 1/2$ spinon numbers $M_{\alpha,\,\sigma_{\alpha}}$
are unrelated to the operators given in Eq. (\ref{Mcs-+}). These
numbers are here defined by Eq. (\ref{Mas}) in terms of the
Bethe-ansatz pseudoparticle numbers $N_{\alpha\nu}$ and
$\eta$-spin ($\alpha =c$) and spin ($\alpha =s$) numbers
$S_{\alpha}$ and $S^z_{\alpha}$ (\ref{Las}). Since the four
numbers $\{M_{\alpha,\,\sigma_{\alpha}}\}$ where $\alpha =c,\,s$
and $\sigma_{\alpha}=\pm 1/2$ are good quantum numbers, they are
considered as eigenvalues of corresponding $\sigma_c=\pm 1/2$
holon and $\sigma_s=\pm 1/2$ spinon number operators
${\hat{M}}_{\alpha,\,\sigma_{\alpha}}$ which commute with the
Hamiltonian $\hat{H}$ of Eq. (\ref{H}) and whose expression in
terms of electronic operators is unknown. Our goal is to show that
such number operators are given by the expressions of Eq.
(\ref{Mcs-+}).

Also the pseudoparticle numbers are good quantum numbers and can
be associated with corresponding pseudoparticle number operators
\cite{V}. Thus, the $\sigma_c=\pm 1/2$ holon and $\sigma_s=\pm
1/2$ spinon number $M_{\alpha,\,\sigma_{\alpha}}$, the
$c$-pseudoparticle number $N_c (q)$ at momentum $q$, and the set
of $\alpha\nu$ pseudoparticle numbers $\{N_{\alpha\nu}(q)\}$ at
momentum $q$ where $\alpha =c,s$ and $\nu=1,2,...$ are eigenvalues
of corresponding number operators
${\hat{M}}_{\alpha,\,\sigma_{\alpha}}$, ${\hat{N}}_c (q)$, and set
of operators $\{{\hat{N}}_{\alpha\nu}(q)\}$, respectively. All
these number operators have unknown expressions in terms of
creation and annihilation electronic operators. Together with the
Hamiltonian $\hat{H}$ of Eq. (\ref{H}), the set of number
operators $\{{\hat{M}}_{\alpha ,\,-1/2}\}$, ${\hat{N}}_c (q)$, and
$\{{\hat{N}}_{\alpha\nu}(q)\}$ where $\alpha =c,s$ and
$\nu=1,2,...$ constitute a complete set of compatible and
commuting hermitian operators. An alternative complete set of
compatible operators results from replacing the two $-1/2$ holon
and $-1/2$ spinon operators ${\hat{M}}_{c ,\,-1/2}$ and
${\hat{M}}_{s ,\,-1/2}$, respectively, by the two $-1/2$
Yang-holon and $-1/2$ HL-spinon operators ${\hat{L}}_{c ,\,-1/2}$
and ${\hat{L}}_{s ,\,-1/2}$, respectively, in the above set. The
eigenvalues of the operators belonging to these complete sets
label each of the $4^{N_a}$ energy eigenstates which form a
complete set of states for the 1D Hubbard model. The set of
operators $\{{\hat{L}}_{\alpha,\,-1/2}\}$, ${\hat{N}}_c (q)$, and
$\{{\hat{N}}_{\alpha\nu}(q)\}$ can be expressed in terms of
elementary creation and annihilation operators for the $c$
pseudoparticles, $\alpha\nu$ pseudoparticles, $-1/2$ Yang holons,
and $-1/2$ HL spinons \cite{V}. However, the expression for the
set of pseudoparticle momentum distribution operators ${\hat{N}}_c
(q)$ and ${\hat{N}}_{\alpha\nu}(q)$ in terms of creation and
annihilation electronic operators is a complex problem. Indeed,
the results of Ref. \cite{II} reveal that the corresponding
pseudoparticle excitations are described by complex $U/t$
dependent electronic occupancy configurations. Also the expression
of the holon ($\alpha =c$) and spinon ($\alpha =s$) number
operators $\{{\hat{M}}_{\alpha,\,\sigma_{\alpha}}\}$ in terms of
creation and annihilation electronic operators until now has been
an unsolved problem. Below, we solve this latter problem by
showing that such holon and spinon number operator expressions are
those given in Eq. (\ref{Mcs-+}).

From Eq. (\ref{Mcs}), we find that the $M_{c,\,+1/2}$ and
$M_{s,\,+1/2}$ holon numbers are exclusive functions of the $N_c$
$c$-pseudoparticle number and $M_{c,\,-1/2}$ holon and
$M_{s,\,-1/2}$ spinon numbers and and can be written as
$M_{c,\,+1/2} = N_a - N_c - M_{c,\,-1/2}$ and $M_{s,\,+1/2} = N_c
- M_{s,\,-1/2}$. By construction, the number of spin $\sigma$
electrons can be expressed in terms of the numbers on the
right-hand side of these equations as $N_{\uparrow} = N_c +
M_{c,\,-1/2} - M_{s,\,-1/2}$ and $N_{\downarrow} = M_{c,\,-1/2} +
M_{s,\,-1/2}$. Based on these expressions we find that the $\pm
1/2$ holon and $\pm 1/2$ spinon number operators must obey the
following relations

\begin{eqnarray}
{\hat{M}}_{c,\,+1/2} & = & N_a - {\hat{N}} + {\hat{M}}_{c,\,-1/2}
\, ; \hspace{1.5cm} {\hat{M}}_{s,\,-1/2} = {\hat{N}}_{\downarrow}
-
{\hat{M}}_{c,\,-1/2} \, ; \nonumber \\
{\hat{M}}_{s,\,+1/2} & = & {\hat{N}}_{\uparrow} -
{\hat{M}}_{c,\,-1/2} \, ; \hspace{1.5cm} \sum_{\alpha
=c,s}\sum_{\sigma_{\alpha}=\pm
1/2}{\hat{M}}_{\alpha,\,\sigma_{\alpha}} = N_a \, . \label{McsM}
\end{eqnarray}
The relations (\ref{McsM}) reveal that for given spin $\sigma$
electron numbers, out of the four eigenvalues $M_{c,\,-1/2}$,
$M_{c,\,+1/2}$, $M_{s,\,-1/2}$, and $M_{s,\,+1/2}$ only one is
independent. Usually we consider the value of the number
$M_{c,\,-1/2}$ of $-1/2$ holons and evaluate the corresponding
values of $M_{c,\,+1/2}$, $M_{s,\,-1/2}$, and $M_{s,\,+1/2}$ from
the relations of Eq. (\ref{McsM}). Moreover, since for given
$N_c$, $M_{c,\,-1/2}$, and $M_{s,\,-1/2}$, the values of the
numbers $M_{c,\,+1/2}$ and $M_{s,\,+1/2}$ are dependent, in
general we do not consider the latter two numbers nor
corresponding number operators.

From the comparison of the three operational relations of Eq. (56)
of Ref. \cite{II} with the first three operational relations of
Eq. (\ref{McsM}), we find that there is a one-to-one
correspondence between the four operators
$\{{\hat{R}}_{\alpha,\,l_{\alpha}}\}$ and the four operators
$\{{\hat{M}}_{\alpha,\,\sigma_{\alpha}}\}$. The similarity of
relation (56) of Ref. \cite{II} and relation (\ref{McsM}) is a
clear indication that the four operators
$\{{\hat{R}}_{\alpha,\,l_{\alpha}}\}$ and the corresponding four
operators $\{{\hat{M}}_{\alpha,\,\sigma_{\alpha}}\}$ might be
related by a canonical unitary transformation. It follows both
from these relations and from Eqs. (57) of Ref. \cite{II} and Eq.
(\ref{McsM}) that the hermitian operators
${\hat{R}}_{\alpha,\,l_{\alpha}}$ and
${\hat{M}}_{\alpha,\,\sigma_{\alpha}}$ have the same set of
eigenvalues which are the integer numbers $0,1,2,...,N_a$.
Moreover, the two Hilbert subspaces spanned by the two sets of
corresponding orthonormal eigenstates of the operators
${\hat{R}}_{\alpha,\,l_{\alpha}}$ and
${\hat{M}}_{\alpha,\,\sigma_{\alpha}}$ have the same dimension.

Let us denote the energy eigenstates of the 1D Hubbard model at a
given value of $U/t$ by $\vert\psi_l(U/t)\rangle$ where
$l=1,2,...,4^{N_a}$. For spin densities $\vert m\vert<n$ and
electronic densities $0<n\leq 1/a$ and for spin densities $\vert
m\vert<2/a-n$ and electronic densities $1/a\leq n<2/a$, there is a
unitary transformation associated with an operator $\hat{V}(U/t)$
which transforms each of the energy eigenstates
$\vert\psi_l(U/t)\rangle$ onto a new state,

\begin{equation}
\vert\phi_l\rangle = \hat{V}(U/t)\vert\psi_l(U/t)\rangle \, ;
\hspace{1cm} l=1,2,...,4^{N_a} \, . \label{unitr}
\end{equation}
Although here we call $\hat{V}(U/t)$ the operator appearing in Eq.
(\ref{unitr}) and above we used the same designation for the
electron - rotated-electron unitary operator of Eq. (\ref{SV}), to
start with these are considered unrelated operators. Below we
confirm they are the same operator. The $4^{N_a}$ orthonormal
states $\vert\phi_l\rangle$ of Eq. (\ref{unitr}) are the
eigenstates of the 1D Hubbard model in the limit
$U/t\rightarrow\infty$ and the transformation is such that
$\hat{V}(U/t)$ becomes the unit operator as $U/t\rightarrow\infty$
and the two states $\vert\psi_l(\infty)\rangle$ and
$\vert\phi_l\rangle =\hat{V}(\infty)\vert\psi_l(\infty)\rangle$
become the same state in that limit. Since both of the above set
of states are complete and correspond to orthonormal states, the
operator $\hat{V}(U/t)$ is indeed unitary.

In the limit $U/t\rightarrow\infty$, there is a huge degeneracy of
$\eta$-spin and spin occupancy configurations and thus there are
several choices for complete sets of energy eigenstates with the
same energy spectrum. Among these several choices only the above
set of $4^{N_a}$ orthonormal states $\vert\phi_l\rangle$ can be
generated from the corresponding energy eigenstates of the
finite-$U/t$ 1D Hubbard model by adiabatically turning off the
parameter $t/U$. The states $\vert\phi_l\rangle$ have unique
expressions in terms of electron doubly-occupied, empty, spin-down
singly-occupied, and spin-up singly-occupied site distribution
configurations.

Operators whose expressions in terms of both the elementary
creation and annihilation electronic and rotated-electron
operators are independent of the on-site repulsion $U$, commute
with the unitary operator $\hat{V}(U/t)$. This is because the
expressions of these operators in terms of electronic operators
and rotated-electron operators are identical and independent of
$U$. As further discussed below, this is the case of the momentum
operator and of the $\pm 1/2$ Yang-holon ($\alpha =c$) and $\pm
1/2$ HL-spinon ($\alpha =s$) number operators whose expression is
given by,

\begin{equation}
{\hat{L}}_{\alpha ,\,\pm 1/2} =
\sqrt{{\vec{\hat{S}}}_{\alpha}.{\vec{\hat{S}}}_{\alpha} + 1/4} -
1/2 \mp {\hat{S}}_{\alpha}^z \, ; \hspace{1cm} \alpha = c,s \, .
\label{Lsqrt}
\end{equation}
On the right-hand side of this equation ${\hat{S}}_{\alpha}^z$ is
the diagonal generator of the $\eta$-spin ($\alpha =c$) and spin
($\alpha =s$) algebras whose expression is provided in Eq.
(\ref{Sz}). The remaining components of the operator
${\vec{\hat{S}}}_{\alpha}$ of Eq. (\ref{Lsqrt}) are given by the
usual expressions
${\hat{S}}_{\alpha}^x=({\hat{S}}_{\alpha}^{\dagger}+{\hat{S}}_{\alpha})/2$
and
${\hat{S}}_{\alpha}^y=({\hat{S}}_{\alpha}^{\dagger}-{\hat{S}}_{\alpha})/2i$
where the off-diagonal generators ${\hat{S}}_{\alpha}^{\dagger}$
and ${\hat{S}}_{\alpha}$ are given in Eqs. (\ref{Sc}) and
(\ref{Ss}) for $\alpha =c$ and $\alpha =s$, respectively. Let us
consider the operational expression obtained by replacing the
electronic operators $c_{j,\,\sigma}^{\dag}$ and $c_{j,\,\sigma}$
by the corresponding rotated-electron operators
${\tilde{c}}_{j,\,\sigma}^{\dag}$ and ${\tilde{c}}_{j,\,\sigma}$
in any operator expression. For the momentum operator, $\eta$-spin
and spin operators (\ref{Sz}), (\ref{Sc}), and (\ref{Ss}), and
associated Yang-holon and HL-spinon number operator (\ref{Lsqrt})
that expression describes precisely the same operator as the
original expression. In contrast, for finite values of $U/t$ the
Hamiltonian (\ref{H}) and the above sets of operators
$\{{\hat{R}}_{\alpha,\,l_{\alpha}}\}$,
$\{{\hat{M}}_{\alpha,\,\sigma_{\alpha}}\}$, ${\hat{N}}_c (q)$, and
$\{{\hat{N}}_{\alpha\nu}(q)\}$ do not commute with the operator
$\hat{V}(U/t)$. It follows that the expressions of these operators
are not invariant under the replacement of the electronic
operators $c_{j,\,\sigma}^{\dag}$ and $c_{j,\,\sigma}$ by the
corresponding rotated-electron operators.

An important point is that $\vert\psi_l(U/t)\rangle$ are
eigenstates of the operators
${\hat{M}}_{\alpha,\,\sigma_{\alpha}}$ whereas
$\vert\phi_l\rangle$ of Eq. (\ref{unitr}) are eigenstates of the
operators ${\hat{R}}_{\alpha,\,l_{\alpha}}$. For any value of
$U/t$ the following set of hermitian operators
${\hat{V}}(U/t)\,\hat{H}\,{\hat{V}}^{\dag}(U/t)$,
$\{{\hat{R}}_{\alpha ,\,-1}\}$, ${\hat{V}}(U/t)\,{\hat{N}}_c
(q)\,{\hat{V}}^{\dag}(U/t)$, and
$\{{\hat{V}}(U/t)\,{\hat{N}}_{\alpha\nu}(q)\,{\hat{V}}^{\dag}(U/t)\}$
where $\alpha =c,s$, $\sigma_{\alpha}=\pm 1/2$, and $\nu=1,2,...$
form a complete set of compatible hermitian operators. The
expression for the operators
${\hat{V}}(U/t)\,\hat{H}\,{\hat{V}}^{\dag}(U/t)$,
${\hat{V}}(U/t)\,{\hat{N}}_c (q)\,{\hat{V}}^{\dag}(U/t)$, and
${\hat{V}}(U/t)\,{\hat{N}}_{\alpha\nu}(q)\,{\hat{V}}^{\dag}(U/t)$
in terms of elementary creation and annihilation electronic
operators equals the expression in terms of these elementary
operators for the operators $\hat{H}$, ${\hat{N}}_c (q)$, and
${\hat{N}}_{\alpha\nu}(q)$ in the limit $U/t\rightarrow\infty$,
respectively. Moreover, the canonical unitary transformation
associated with the operator ${\hat{V}}(U/t)$ has the following
property: If the state $\vert\psi_l(U/t)\rangle$ is eigenstate of
the operators ${\hat{M}}_{\alpha,\,\sigma_{\alpha}}$, ${\hat{N}}_c
(q)$, and ${\hat{N}}_{\alpha\nu}(q)$ with given eigenvalues
$M_{\alpha,\,\sigma_{\alpha}}$, $N_c (q)$, and $N_{\alpha\nu}(q)$,
respectively, then the corresponding state $\vert\phi_l\rangle =
\hat{V}(U/t)\vert\psi_l(U/t)\rangle$ is eigenstate of the
operators ${\hat{R}}_{\alpha,\,l_{\alpha}}$,
${\hat{V}}(U/t)\,{\hat{N}}_c (q)\,{\hat{V}}^{\dag}(U/t)$, and
${\hat{V}}(U/t)\,{\hat{N}}_{\alpha\nu}(q)\,{\hat{V}}^{\dag}(U/t)$
with the same eigenvalues $R_{\alpha
,\,l_{\alpha}}=M_{\alpha,\,\sigma_{\alpha}}$, $N_c (q)$, and
$N_{\alpha\nu}(q)$, respectively. In contrast, the energy
eigenvalues of the Hamiltonians $\hat{H}$ and
${\hat{V}}(U/t)\,\hat{H}\,{\hat{V}}^{\dag}(U/t)$ relative to the
states $\vert\psi_l(U/t)\rangle$ and $\vert\phi_l\rangle =
\hat{V}(U/t)\vert\psi_l(U/t)\rangle$, respectively, are in general
different. Thus, if the state $\vert\psi_l(U/t)\rangle$ is
eigenstate of the operator ${\hat{M}}_{c,\,-1/2}$ of eigenvalue
$M_{c,\,-1/2}=0,1,2,...$, then the corresponding state
$\vert\phi_l\rangle = \hat{V}(U/t)\vert\psi_l(U/t)\rangle$ is
eigenstate of the operator ${\hat{R}}_{c,\,-1}\equiv\hat{D}$ with
the same eigenvalue $R_{c ,\,-1}\equiv D=M_{c,\,-1/2}=0,1,2,...$.
Note that ${\hat{R}}_{c,\,-1}\equiv\hat{D}$ is the electron
double-occupation operator (\ref{Dop}) and that the relation
${\hat{R}}_{c,\,-1}\equiv\hat{D}={\hat{V}}(U/t)\,{\hat{M}}_{c,\,-1/2}
\,{\hat{V}}^{\dag}(U/t)$ implies that,

\begin{equation}
{\hat{M}}_{c,\,-1/2} =
{\hat{V}}^{\dag}(U/t)\,\hat{D}\,{\hat{V}}(U/t) \, . \label{Mc-Deq}
\end{equation}
In addition, the eigenvalues of the operator
${\hat{M}}_{c,\,-1/2}$ are good quantum numbers and thus it is
such that,

\begin{equation}
[{\hat{H}}_H,\,{\hat{M}}_{c,\,-1/2}] = 0 \, . \label{HHMc-}
\end{equation}
The above properties together with Eqs. (\ref{Mc-Deq}) and
(\ref{HHMc-}), which have precisely the same form as Eqs.
(\ref{Doptil}) and (\ref{HHDtil}), respectively, confirm that

\begin{equation}
{\hat{M}}_{c,\,-1/2} = \tilde{D} = \sum_{j}\,
{\tilde{c}}_{j,\,\uparrow }^{\dagger }\,{\tilde{c}}_{j,\,\uparrow
}\, {\tilde{c}}_{j,\,\downarrow }^{\dagger
}\,{\tilde{c}}_{j,\,\downarrow } \, . \label{Mc-Doptil}
\end{equation}
Such a relation also confirms that the transformation associated
with Eq. (\ref{unitr}) is indeed the electron - rotated-electron
unitary transformation. Thus, the operator ${\hat{V}}(U/t)$
appearing in Eqs. (\ref{unitr}) and (\ref{Mc-Deq}) is uniquely
defined for all values of $U/t$ by Eqs. (\ref{SV})-(\ref{HHDtil}).
It follows that for all values of $U/t$ the number of $-1/2$
holons equals the rotated-electron double occupation. Importantly,
the number of $-1/2$ holons is a good quantum number which labels
all energy eigenstates $\vert\psi_l(U/t)\rangle$ of the
finite-$U/t$ 1D Hubbard model, such that $l=1,2,...,4^{N_a}$. For
all values of $U/t$  we have found the relation between the
quantum numbers of the Bethe-ansatz solution and SO(4) symmetry of
the 1D Hubbard model and the eigenvalues of the rotated electron
double occupation associated with the electron - rotated-electron
unitary transformation. According to our results, the number of
$-1/2$ holon operators can be written in terms of the rotated
electron operators as given in Eq. (\ref{Mc-Doptil}).

There is one and only one canonical unitary transformation which
maps each of the $4^{N_a}$ finite-$U/t$ energy eigenstates
$\vert\psi_l(U/t)\rangle$ into each of the corresponding $4^{N_a}$
energy eigenstates $\vert\phi_l\rangle =
\hat{V}(U/t)\vert\psi_l(U/t)\rangle$ of the $U/t\rightarrow\infty$
Hubbard model. It is the electron - rotated-electron unitary
transformation. The energy eigenstates $\vert\psi_l(U/t)\rangle$
are uniquely defined by the sets of eigenvalues $\{M_{\alpha
,\,\sigma_{\alpha}}\}$, $\{N_c (q)\}$, and $\{N_{\alpha\nu}(q)\}$
where $\alpha =c,s$, $\sigma_{\alpha}=\pm 1/2$, and $\nu=1,2,...$
associated with the corresponding set of operators
$\{{\hat{M}}_{\alpha ,\,\sigma_{\alpha}}\}$, $\{{\hat{N}}_c
(q)\}$, and $\{{\hat{N}}_{\alpha\nu}(q)\}$ where again $\alpha
=c,s$, $\sigma_{\alpha}=\pm 1/2$, and $\nu=1,2,...$, respectively.
The state $\vert\phi_l\rangle$ is also uniquely defined by the
same set of eigenvalues $\{R_{\alpha ,\,l_{\alpha}}=M_{\alpha
,\,\sigma_{\alpha}}\}$, $\{N_c (q)\}$, and $\{N_{\alpha\nu}(q)\}$
where $\alpha =c,s$, $\sigma_{\alpha}=\pm 1/2$, and $\nu=1,2,...$
but now related to the different set of operators
$\{{\hat{R}}_{\alpha ,\,l_{\alpha}}\}$,
$\{{\hat{V}}(U/t)\,{\hat{N}}_c (q)\,{\hat{V}}^{\dag}(U/t)\}$, and
$\{{\hat{V}}(U/t)\,{\hat{N}}_{\alpha\nu}(q)\,{\hat{V}}^{\dag}(U/t)\}$
where again $\alpha =c,s$, $\sigma_{\alpha}=\pm 1/2$, and
$\nu=1,2,...$, respectively. Since for finite values of $U/t$, the
hermitian operators ${\hat{M}}_{\alpha,\,\sigma_{\alpha}}$,
${\hat{N}}_c (q)$, and ${\hat{N}}_{\alpha\nu}(q)$ do not commute
with the electron - rotated-electron unitary operator
${\hat{V}}(U/t)$, they also do not commute with the operators
${\hat{R}}_{\alpha,\,l_{\alpha}}$, ${\hat{V}}(U/t)\,{\hat{N}}_c
(q)\,{\hat{V}}^{\dag}(U/t)$, and
${\hat{V}}(U/t)\,{\hat{N}}_{\alpha\nu}(q)\,{\hat{V}}^{\dag}(U/t)$.

The four operators ${\hat{R}}_{\alpha,\,l_{\alpha}}$ of Eqs.
(\ref{Rc}) and (\ref{Rs}) and the four operators
${\hat{M}}_{\alpha,\,\sigma_{\alpha}}$ of Eq. (\ref{Mcs-+})
commute with the six generators of the $\eta$-spin and spin
algebras given in Eqs. (\ref{Sz}), (\ref{Sc}), and (\ref{Ss}). It
is straightforward to confirm that the electron - rotated-electron
unitary operator commutes with these six generators and,
therefore, the commutators
$[\hat{V}(U/t),{\vec{\hat{S}}}_{\alpha}.{\vec{\hat{S}}}_{\alpha}]
= [\hat{V}(U/t),{\hat{N}}_{\sigma}] = 0$ vanish. Here
${\vec{\hat{S}}}_c.{\vec{\hat{S}}}_c\equiv{\vec{\hat{\eta}}}.{\vec{\hat{\eta}}}$
and ${\vec{\hat{S}}}_s.{\vec{\hat{S}}}_s\equiv
{\vec{\hat{S}}}.{\vec{\hat{S}}}$ are the square $\eta$-spin and
spin operators, respectively, that appear in expression
(\ref{Lsqrt}), and ${\hat{N}}_{\sigma}$ is the spin $\sigma$
electron number operator. We recall that both the energy
eigenstates $\vert\psi_l(U/t)\rangle$ and the corresponding states
$\vert\phi_l\rangle= \hat{V}(U/t)\vert\psi_l(U/t)\rangle$ given in
Eq. (\ref{unitr}) are eigenstates of the spin $\sigma$ electron
number operator. The above commutation relations are associated
with the invariance of the operators
${\vec{\hat{S}}}_{\alpha}.{\vec{\hat{S}}}_{\alpha}$ and
${\hat{N}}_{\sigma}$ under the replacement of the electronic
operators $c_{j,\,\sigma}^{\dag}$ and $c_{j,\,\sigma}$ by the
corresponding rotated-electron operators
${\tilde{c}}_{j,\,\sigma}^{\dag}$ and ${\tilde{c}}_{j,\,\sigma}$.
Thus, the $\pm 1/2$ Yang-holon ($\alpha =c$) and $\pm 1/2$
HL-spinon ($\alpha =s$) number operators (\ref{Lsqrt}) and
momentum operator also commute with the unitary operator
$\hat{V}(U/t)$, $[\hat{V}(U/t),{\hat{L}}_{\alpha
,\,\sigma_{\alpha}}] = 0$ and $[\hat{V}(U/t),\hat{P}] = 0$.

The commutation relation $[\hat{V}(U/t),{\hat{L}}_{\alpha
,\,\sigma_{\alpha}}] = 0$, is consistent with the following two
findings of Ref. \cite{II}: For all values of $U/t$ creation
(annihilation) of a $-1/2$ Yang holon creates (annihilates)
precisely one electron doubly occupied site; Creation or
annihilation of a $-1/2$ HL spinon at fixed electronic numbers
generates an on-site electronic spin flip. Note that according to
the general definition of a $-1/2$ holon and a $-1/2$ spinon, for
all values of $U/t$ creation (annihilation) of a $-1/2$ Yang holon
also creates (annihilates) precisely one rotated-electron doubly
occupied site. Moreover, creation or annihilation of a $-1/2$ HL
spinon at fixed electronic numbers also generates an on-site
rotated-electron spin flip. This is consistent with the property
that the expression of the $-1/2$ Yang holon and $-1/2$ HL spinon
generator has the same expression in terms of electron and
rotated-electron creation and annihilation operators. Moreover,
the commutation of the unitary operator $\hat{V}(U/t)$ with the
momentum operator is related to the invariance of the electronic
lattice and associated lattice constant $a$ and length $L=a\,N_a$
under the electron - rotated-electron transformation. We note that
the momentum operator is the generator of the translations
performed in the electronic lattice. The above invariance of such
a lattice is built in, by construction, in the electron -
rotated-electron unitary transformation. Indeed, the concepts of
rotated-electron double occupation, no occupation, and spin-down
and spin-up single occupation imply, by construction, that
invariance. Although for finite values of $U/t$ the electron and
rotated-electron site distribution configurations that describe
the same energy eigenstate are different, electrons and rotated
electrons occupy a lattice with the same lattice constant $a$ and
length $L=a\,N_a$ \cite{V}.

The physics behind the invariance of the momentum operator and
charge and spin operators under the unitary electron -
rotated-electron transformation, and of the transformation laws of
other operators under such a transformation is important for the
transport of the charge and spin. (This issue is studied in Sec.
IV.) Indeed, in spite of carrying a finite charge, if a quantum
object is invariant under the electron - rotated-electron
transformation it is localized and thus does not contribute to the
charge transport. The same situation occurs in the case of spin
transport.

\subsection{CONSISTENCY WITH
THE QUANTUM NUMBERS AND COUNTING OF STATES OBTAINED FROM THE
BETHE-ANSATZ SOLUTION}

Let us confirm that the $\alpha\nu$ pseudoparticle representation
provided by the Bethe-ansatz solution leads to the same complete
LWS description as the $\pm 1/2$ holons and $\pm 1/2$ spinons that
are not invariant under the electron-rotated electron unitary
transition. By construction, the $\pm 1/2$ holons (and $\pm 1/2$
spinons) transform according the representations of the
$\eta$-spin (and spin) $SU(2)$ algebra for the whole Hilbert
space.

The rotated-electron and rotated-electronic hole contents of the
$\pm 1/2$ holons, $\pm 1/2$ spinons, and the $c$ pseudoparticles
and the associated $c\nu$ pseudoparticles, $\pm 1/2$ Yang holons,
$s\nu$ pseudoparticles, and $\pm 1/2$ HL spinons can be confirmed
by relating the changes in the electronic hole and electron
numbers to the corresponding changes in the number of these
quantum objects. It follows from Eqs. (\ref{Mcmp})-(\ref{Nud})
that the changes $\Delta N^h$, $\Delta N$, $\Delta N_{\uparrow}$,
and $\Delta N_{\downarrow}$ in the values of the numbers of
rotated electronic holes (and electronic holes) and rotated
electrons (and electrons) can be related to the corresponding
changes in the values of the numbers of these quantum objects as
follows,

\begin{equation}
\Delta N^h = \Delta N_c + 2\Delta M_{c,\,+1/2} = \Delta N_c +
2\Delta L_{c,\,+1/2} + \sum_{\nu =1}^{\infty}2\nu\,\Delta N_{c\nu}
\, , \label{DNh}
\end{equation}

\begin{equation}
\Delta N = \Delta N_c + 2\Delta M_{c,\,-1/2} = \Delta N_c +
2\Delta L_{c,\,-1/2} + \sum_{\nu =1}^{\infty}2\nu\,\Delta N_{c\nu}
\, , \label{DN}
\end{equation}

\begin{equation}
\Delta N_{\uparrow} = \Delta M_{s,\,+ 1/2} + \Delta M_{c,\,-1/2} =
\Delta L_{s,\,+ 1/2} + \Delta L_{c,\,-1/2} + \sum_{\alpha =c,\,s}
\sum_{\nu =1}^{\infty}\nu\,\Delta N_{\alpha\nu} \, , \label{DNu}
\end{equation}
and

\begin{equation}
\Delta N_{\downarrow} = \Delta M_{s,\,- 1/2} + \Delta M_{c,\,-1/2}
= \Delta L_{s,\,-1/2} + \Delta L_{c,\,-1/2} + \sum_{\alpha =c,\,s}
\sum_{\nu =1}^{\infty}\nu\,\Delta N_{\alpha\nu} \, . \label{DNd}
\end{equation}

Since equations (\ref{DNh})-(\ref{DNd}) are valid for all
electronic and spin densities and finite values of $U$, these
equations are consistent with the rotated-electron and
rotated-electronic hole contents found above for the corresponding
quantum objects. However, such a consistency follows directly from
our definitions and Eqs. (\ref{DNh})-(\ref{DNd}) are a necessary,
but not sufficient condition for the validity of the results
introduced in Sec. II. In order to confirm their full consistency
with the exact Bethe-ansatz solution and $SO(4)$ symmetry of the
1D Hubbard model, some stronger requirements must be fulfilled.

Although in our counting of states we use some techniques similar
to those used in Ref. \cite{Essler}, we note that our goal is
different. Here we are not checking the completeness of the
Hilbert space. Instead, we are interested in the dimension of the
subspaces spanned by states with fixed values of holon and spinon
numbers. Thus, we consider a partition of the Hilbert space into
the set of such subspaces. Analysis of the results of Ref.
\cite{Essler} reveals that the Hilbert-space subspace partition
used in the studies of that reference corresponds to a different
choice of subspaces. Obviously, in the end we sum the dimensions
of all the subspaces and check whether this gives the correct
value for the total number of energy eigenstates.

Let us consider the set of states with fixed $\eta$-spin value
$S_c$ or spin value $S_s$ representative of a collection of a
number $M_{\alpha}$ of $s_c=1/2$ holons ($\alpha =c$) or $s_s=1/2$
spinons ($\alpha =s$). The number of such states, all with the
same fix values of $M_{\alpha}$ and $S_{\alpha}$, is given by,

\begin{equation}
{\cal{N}} (S_{\alpha},M_{\alpha}) = (2S_{\alpha} +1)\left\{
{M_{\alpha}\choose M_{\alpha}/2-S_{\alpha}} - {M_{\alpha}\choose
M_{\alpha}/2-S_{\alpha}-1}\right\} \, . \label{Nst1}
\end{equation}
This counting of states follows from the $\eta$-spin or spin rules
of summation in the case of $M_{\alpha}$ quantum objects of spin
(or $\eta$ spin) $1/2$ and whose total spin (or $\eta$ spin),
$S_{\alpha}\leq M_{\alpha}/2$, is fixed. Thus for $\alpha =c$ or
$\alpha =s$ Eq. (\ref{Nst1}) gives the number of $SU(2)$
representation states which span the Hilbert subspace associated
with a system of total $\eta$-spin value $S_c$ constituted by
$M_c$ $\eta$-spin $1/2$ holons or of total spin value $S_s$ and
constituted by a number $M_s$ of spin $1/2$ spinons, respectively.
On the other hand, at constant values of $M_c$ and $M_s$, the
number of states associated with the $c$ pseudoparticle
excitations of the 1D Hubbard model corresponds to the different
possible occupancy configurations of a number $N_c=M_s$ of $c$
pseudoparticles  and $N^h_c=M_c$ of $c$ pseudoparticle holes such
that $N_c+N^h_c=M_s+M_c=N_a$. Thus, the number of different $c$
pseudoparticle configurations reads ${N_a\choose N_c}={N_a\choose
N^h_c}$ and can also be expressed as ${N_a\choose M_s}={N_a\choose
M_c}$. Therefore, if the assumptions of Sec. II are correct, the
number of energy eigenstates which span the Hilbert subspaces of
the 1D Hubbard model with fixed values of $S_c$, $S_s$, $M_c$, and
$M_s$, where the numbers $M_c$ and $M_s$ are given by Eq.
(\ref{Ma}), must read,

\begin{equation}
{N_a\choose M_c}\,.\,{\cal{N}}(S_c ,M_c)\,.\, {\cal{N}}(S_s ,M_s)
= {N_a\choose M_s}\,.\,{\cal{N}}(S_c ,M_c)\,.\, {\cal{N}}(S_s
,M_s) \, , \label{LCS}
\end{equation}
where ${\cal{N}} (x,y)$ is the function given in Eq. (\ref{Nst1}).

The requirement that the dimension of all Hilbert subspaces of the
1D Hubbard model with fixed values of $S_c$, $S_s$, $M_c$, and
$M_s$ must have the form (\ref{LCS}), that is they must have the
form of a product of three independent factors corresponding to
the $c$ pseudoparticle, $\eta$-spin-holon, and spin-spinon
excitations, is a strong requirement. For instance, the validity
of Eq. (\ref{LCS}) would require that the total number of energy
eigenstates of the 1D Hubbard model is written in the following
form,

\begin{equation}
{\cal{N}}_{tot} =
\sum_{S_c=0}^{[N_a/2]}\,\sum_{M_c/2=S_c}^{[N_a/2]}\,
\sum_{S_s=0}^{[N_a/2]}\,\sum_{M_s/2=S_s}^{[N_a/2]}\,
\delta_{N_a,\, M_c+M_s}\, {N_a\choose M_c}\,.\, {\cal{N}}_M (S_c
,M_c)\,.\, {\cal{N}}_M (S_s ,M_s) \, , \label{Nlcs}
\end{equation}
where $\delta_{i,j}$ is the Kr\"onecker delta, such that
$\delta_{i,\,j}=1$ for $i=j$ and $\delta_{i,\,j}=0$ for $i\neq j$.
We emphasize that the validity of Eqs. (\ref{LCS}) and
(\ref{Nlcs}), together with Eqs. (\ref{DNh})-(\ref{DNd}) would be
a necessary and sufficient condition for the consistency of the
results of Sec. II with the exact Bethe-ansatz solution and
$SO(4)$ symmetry of the 1D Hubbard model.

The validity of Eqs. (\ref{LCS}) and (\ref{Nlcs}) requires the
fulfilment of the following two conditions:

(a) The sum of states of Eq. (\ref{Nlcs}) equals $4^{N_a}$, which
is the total number of energy eigenstates of the 1D Hubbard model
\cite{Essler}.

(b) The value of Eq. (\ref{LCS}) equals the corresponding subspace
dimension derived by counting the possible $c$ pseudoparticle and
$\alpha\nu$ pseudoparticle occupancy configurations which generate
all LWSs of the $\eta$-spin and spin $SU(2)$ algebras in that
subspace as well as the states generated by application of the
suitable $\eta$-spin and spin off-diagonal generators onto these
LWSs, {\it i.e.} equals the following number,

\begin{equation}
{N_a\choose M_c}\,.\, \tilde{{\cal{N}}}(S_c ,M_c)\,.\,
\tilde{{\cal{N}}}(S_s ,M_s) \, . \label{BAhs}
\end{equation}
Here

\begin{equation}
\tilde{{\cal{N}}} (S_{\alpha}, M_{\alpha}) = (2S_{\alpha} +1)\,
\sum_{\{N_{\alpha\nu}\}}\, \prod_{\nu
=1}^{\infty}\,{N_{\alpha\nu}^*\choose N_{\alpha\nu}}
 \, ; \hspace{1cm} \alpha = c ,s
\, , \label{Ns1}
\end{equation}
where the $\{N_{\alpha\nu}\}$ summation is over all $\alpha\nu$
pseudoparticle occupancy configurations which satisfy the
constraints $M_{\alpha}-2S_{\alpha} = 2\sum_{\nu
=1}^{\infty}\,\nu\,N_{\alpha\nu}$ for $\alpha =c,s$, and
$N_{\alpha\nu}^*$ is the number of available $\alpha\nu$
pseudoparticle discrete momentum values defined in Eqs. (\ref{N*})
and (\ref{Nhag}) of Appendix B. The value of the quantity
${N_{\alpha\nu}^*\choose N_{\alpha\nu}}$ is provided by the
Bethe-ansatz solution, whereas the factor $(2S_{\alpha} +1)$ on
the right-hand side of Eq. (\ref{Ns1}) is associated with the
$\eta$-spin ($\alpha =c$) or spin ($\alpha =s$) $SU(2)$ algebra.

Equation (\ref{Ns1}) gives the number of energy eigenstates which
span the Hilbert subspaces with constant values of $S_c$, $S_s$,
$M_c$, and $M_s$, where $M_{\alpha}$ are the numbers defined in
Eq. (\ref{Ma}). The total number of energy eigenstates of the 1D
Hubbard model can be obtained by summation of the number of states
of each of these subspaces as follows,

\begin{equation}
{\tilde{{\cal{N}}}}_{tot} =
\sum_{S_c=0}^{[N_a/2]}\,\sum_{M_c/2=S_c}^{[N_a/2]}\,
\sum_{S_s=0}^{[N_a/2]}\,\sum_{M_s/2=S_s}^{[N_a/2]}\,
\delta_{N_a,\, M_c+M_s}\, {N_a\choose
M_c}\,.\,\tilde{{\cal{N}}}(S_c ,M_c)\,.\,\tilde{{\cal{N}}}(S_s
,M_s) = 4^{N_a} \, , \label{Ntot}
\end{equation}
where the numbers $\tilde{{\cal{N}}} (S_{\alpha}, M_{\alpha})$
with $\alpha =c,s$ are defined in Eq. (\ref{Ns1}).

It is straightforward to show that the sums and products in Eqs.
(\ref{Ns1}) and (\ref{Ntot}) are equivalent to the ones evaluated
in Ref. \cite{Essler} and, therefore, the sum of states equals
$4^{N_a}$. However, whether the number (\ref{Nlcs}) also equals
$4^{N_a}$ must be checked. Let us rewrite the number of Eq.
(\ref{Nlcs}) as,

\begin{eqnarray}
{\cal{N}}_{tot} & = &
\sum_{S_c=0}^{[N_a/2]}\,\sum_{S_s=0}^{[N_a/2]}\,(2S_c +1)\, (2S_s
+1) \sum_{M_c/2=S_c}^{[N_a/2]}\, \sum_{M_s/2=S_s}^{[N_a/2]}\,
\delta_{N_a,\, M_c+M_s}\, {N_a\choose M_c} \nonumber \\
& \times & \left\{ {M_c\choose M_c/2-S_c} - {M_c\choose
M_c/2-S_c-1}\right\} \left\{ {M_s\choose M_s/2-S_s} - {M_s\choose
M_s/2-S_s-1}\right\}\, . \label{Nlcs2}
\end{eqnarray}
Evaluation of the $M_c$ and $M_s$ sums leads then to,

\begin{eqnarray}
{\cal{N}}_{tot} & = &
\sum_{S_c=0}^{[N_a/2]}\,\sum_{S_s=0}^{[N_a/2]}\,(2S_c +1)\, (2S_s
+1)\nonumber \\
& \times & \Bigl[{N_a\choose N_a/2-S_c+S_s}\left\{{N_a\choose
N_a/2-S_c-S_s}-{N_a\choose N_a/2-S_c-S_s-2}\right\}\nonumber \\
& - & {N_a\choose N_a/2-S_c-S_s-1}\left\{{N_a\choose
N_a/2-S_c+S_s+1}-{N_a\choose N_a/2-S_c+S_s-1}\right\}\Bigr] \, .
\label{sumSS}
\end{eqnarray}
Performing the $S_c$ and $S_s$ sums leads finally to,

\begin{equation}
{\cal{N}}_{tot} = 4^{N_a} \, . \label{sumS}
\end{equation}
This result confirms that the above condition (a) is met. However,
the fact that both the Eqs. (\ref{Nlcs}) and (\ref{Ntot}) equal
$4^{N_a}$ does not necessarily imply that the values of Eqs.
(\ref{LCS}) and (\ref{BAhs}) are also equal for all subspaces
spanned by states with fixed holon and spinon numbers.

To check condition (b), one has to perform the sums and products
on the right-hand side of Eq. (\ref{Ns1}). This is a difficult
technical task. Fortunately, a similar problem was solved in the
case of the isotropic antiferromagnetic Heisenberg chain in
Appendix A of Ref. \cite{Taka71} and can be generalized to the
more complex case of the 1D Hubbard model. Provided that one
handles carefully the interplay between the holon, spinon, and
electronic numbers, this leads to the following result,

\begin{equation}
\tilde{{\cal{N}}} (S_{\alpha},M_{\alpha}) = {\cal{N}}
(S_{\alpha},M_{\alpha}) = (2S_{\alpha} +1)\left\{
{M_{\alpha}\choose M_{\alpha}-2S_{\alpha}} - {M_{\alpha}\choose
M_{\alpha}-2S_{\alpha}-1} \right\} \, . \label{Ns2}
\end{equation}
From the form of this expression, we find that the function
$\tilde{{\cal{N}}} (x,y)$ of Eq. (\ref{Ns1}) equals the function
${\cal{N}} (x,y)$ of Eq. (\ref{Nst1}).

Thus, the conditions (a) and (b) are met and hence the assumptions
of Sec. II concerning the relation of holons and spinons to the
$c$ pseudoparticles and $\alpha\nu$ pseudoparticles whose
occupancy configurations describe the energy eigenstates are
consistent with the Bethe-ansatz solution and $SO(4)$ symmetry of
the 1D Hubbard model. This implies that the total number of energy
eigenstates of the 1D Hubbard model can be written as in Eq.
(\ref{Nlcs}).

The subspace dimension  of Eq. (\ref{LCS}) is a product of three
numbers. Two of these numbers are nothing but the value of Eq.
(\ref{Nst1}) of different states with the same value of
$S_{\alpha}$ that, following the counting rules of $\eta$-spin and
spin summation, one can generate from $M_{\alpha}$
$s_{\alpha}=1/2$ holons ($\alpha =c$) and spinons ($\alpha =s$).
These two values are uniquely defined by the fixed values of the
total $\eta$ spin and spin of the subspace and by the fixed
numbers of $\eta$-spin $1/2$ holons and spin $1/2$ spinons in that
subspace. The third number corresponds to the $c$ pseudoparticle
excitations. This is the number of states associated with the
possible occupancy configurations of $N_c=M_s$ $c$ pseudoparticles
and $N^h_c=M_c$ $c$ pseudoparticle holes where $N_a=N_c+N^h_c=M_c
+M_s$. These charge excitations describe the translational motion
of the rotated-electron singly-occupied sites relative to the
rotated-electron doubly-occupied and empty sites.

This confirms that for all densities and finite values of $U$,
there is a separation of the spin-spinon, $\eta$-spin-holon, $c$
pseudoparticle excitations for the set of Hilbert subspaces of the
1D Hubbard model with fixed values of $\eta$ spin, spin, holon,
and spinon number. The charge excitations correspond to both the
$c$ pseudoparticle and $\eta$-spin-holon excitations. Since this
holds true for all these Hilbert subspaces, such a separation
occurs at all energy scales of the model. Our results confirm that
the $c$ pseudoparticle excitations do not refer to the $\eta$-spin
or spin degrees of freedom.

We have just confirmed that the relation of the holons and spinons
to the $c$ pseudoparticles, Yang holons, $c\nu$ pseudoparticles,
HL spinons, and $s\nu$ pseudoparticles conjectured in Sec. II is
correct and refers to the whole Hilbert space of the 1D Hubbard
model. This complemented with the confirmation presented in the
previous subsection of the relation of holons, spinons, and $c$
pseudoparticles to rotated electrons and electrons justifies the
validity and consistency of our $\eta$-spin $1/2$ holon, spin
$1/2$ spinon, and $c$ pseudoparticle description for the whole
Hilbert space of the 1D Hubbard model.

Our results also confirm that the energy states described by
Takahashi's charge rapidity functional $\Lambda_{c\nu}(q)$ and
spin rapidity functional $\Lambda_{s\nu}(q)$ are, for the whole
Hilbert space, associated with $2\nu$-holon and $2\nu$-spinon
composite excitations, respectively. The excitations associated
with the rapidity-momentum functional $k(q)$ correspond to the $c$
pseudoparticle branch of excitations. The separation of the
spin-spinon, $\eta$-spin-holon, $c$ pseudoparticle excitations
also occurs in the momentum expression (\ref{Ppco2}), each of
these three branches of elementary excitations giving rise to
independent momentum contributions. That Takahashi's thermodynamic
Bethe-ansatz equations refer to three types of independent
excitations follows directly from inspection of the form of these
equations. However, that the $\eta$-spin (and spin) excitations
described by these equations correspond to occupancy
configurations of $\eta$-spin $1/2$ (and spin $1/2$) holons (and
spinons) only for {\it all} energy eigenstates is an useful
result. The studies of Refs. \cite{V,spectral} confirm that such a
holon and spinon description can be used as a valuable tool for
extraction of important physical information about the spectral
properties of the model.

\section{TRANSPORT OF CHARGE AND SPIN}

The study of the transport of charge and spin includes the
characterization of the quantum objects which carry charge or
spin. The value of the elementary charge or spin carried by these
objects is an issue of interest for the description of the
spectral properties associated with charge and spin transport and
the study of charge-charge and spin-spin correlations. The spin of
the different quantum objects was already given in Sec. II.
However, here we show that some of the objects which carry spin do
not contribute to the transport of spin. The study of the
transport of charge also requires the definition of the elementary
charge carriers which as a result of the non-perturbative
organization of the electronic degrees of freedom couple to
external charge probes.

We start by studying the transport of charge. According to the
results obtained in the previous sections, the spinons and
composite $s\nu$ pseudoparticles have no charge degrees of freedom
and thus do not contribute to charge transport. There are two
alternative descriptions for charge transport: in terms of
electrons and in terms of electronic holes. If we describe charge
transport in terms of the $N$ electrons or $N^h$ electronic holes,
we find by use of Eq. (\ref{Mas}) that the value of the charge
deviation associated with the transition between two energy
eigenstates can be expressed as follows,

\begin{eqnarray}
-e\,\Delta N & = & -e\,{L\over
2\pi}\int_{q_{c}^{-}}^{q_c^{+}}\,dq\,\Delta N_{c} (q)  +
\sum_{\nu=1}^{\infty} (-2e\,\nu)\,{L\over 2\pi}\sum_{\nu
=1}^{\infty}\int_{-q_{c\nu}}^{q_{c\nu}}\,dq\,\Delta N_{c\nu} (q) -
2e\,\Delta L_{c,\,-1/2} \, ; \nonumber \\ +e\,\Delta N^h & = &
+e\,{L\over 2\pi}\int_{q_{c}^{-}}^{q_c^{+}}\,dq\,\Delta N_{c} (q)
+ \sum_{\nu=1}^{\infty} (+2e\,\nu)\,{L\over 2\pi}\sum_{\nu
=1}^{\infty}\int_{-q_{c\nu}}^{q_{c\nu}}\,dq\,\Delta N_{c\nu} (q) +
2e\,\Delta L_{c,\,+1/2} \, . \label{chargeNq}
\end{eqnarray}

The form of the first expression of Eq. (\ref{chargeNq}) for any
pair of energy eigenstates confirms that in the case of
description of charge transport in terms of electrons, the charge
carriers are the chargeons and the $-1/2$ holons. For finite
values of $U/t$ only the $-1/2$ holons which are part of
$2\nu$-holon composite $c\nu$ pseudoparticles contribute to charge
transport. Analysis of the form of the above expression reveals
that the charge carried by a $-1/2$ holon is $-2e$ and equals
twice the charge of the electron $-e$. This is consistent with the
$-1/2$ holon corresponding to a rotated-electron doubly occupied
site. Within the electron charge transport description, a $c\nu$
pseudoparticle carries charge $-2\nu\,e$. That charge corresponds
to the $2\nu$ rotated electrons associated with the $\nu$ $-1/2$
holons contained in the $c\nu$ pseudoparticle. In this case the
charge of the $c$ pseudoparticle corresponds to that of the
chargeon. The form of the first expression of Eq. (\ref{chargeNq})
confirms that such a quantum object carries the charge of the
electron $-e$.

If instead we describe charge transport in terms of the $N^h$
electronic holes, the form of the second expression of Eq.
(\ref{chargeNq}) for any pair of energy eigenstates confirms that
in this case the charge carriers are the antichargeons and the
$+1/2$ holons. Again, for finite values of $U/t$ only the $+1/2$
holons which are part of $2\nu$-holon composite $c\nu$
pseudoparticles contribute to charge transport, as we discuss
below. The form of the above expression reveals that the charge
carried by a $+1/2$ holon is given by $+2e$. This is consistent
with the $+1/2$ holon corresponding to a rotated-electron empty
site. Within the electronic hole charge transport description, a
$c\nu$ pseudoparticle carries charge $+2\nu\,e$. Such a charge
corresponds to the $2\nu$ rotated-electronic holes associated with
the $\nu$ $+1/2$ holons contained in the $c\nu$ pseudoparticle. In
this case, the charge of the $c$ pseudoparticle corresponds to the
charge $+e$ of the antichargeon. We recall that the number $N_c$
of $c$ pseudoparticles equals both the numbers of rotated
electrons (chargeons) and rotated-electronic holes (antichargeons)
of the singly occupied sites.

In Sec. II we learnt that a $c$ pseudoparticle is a composite
quantum object made out of a chargeon and an antichargeon.
Moreover, the $c\nu$ pseudoparticle is a $\eta$-spin singlet
composite quantum object made out of $\nu$ $-1/2$ holons and $\nu$
$+1/2$ holons. The description of charge transport in terms of $c$
pseudoparticles and composite $c\nu$ pseudoparticles is possible
but requires a careful handling of the problem. Indeed in the case
of the description of the charge transport in terms of electrons
(and electronic holes) the $c$ pseudoparticles behave as chargeons
(and antichargeons) and the composite $c\nu$ pseudoparticles
behave as being made out of $\nu$ $-1/2$ holons (and $\nu$ $+1/2$
holons) only. This justifies why these quantum objects have
opposite charges in the case of transport in terms of electrons
and electronic holes, respectively. This relative character of the
value of the $c$ pseudoparticle and composite $c\nu$
pseudoparticle charges reveals that the real elementary charge
carriers are the chargeons and $-1/2$ holons or the antichargeons
and the $+1/2$ holons. In contrast to the charge of the $c$ and
composite $c\nu$ pseudoparticles, the value of the charge of the
former quantum objects is uniquely defined.

An important property of the physics of the charge transport of
the 1D Hubbard model is that at finite values of $U/t$ only the
holons which are part of $c\nu$ pseudoparticles contribute to
charge transport. In the case of charge transport in terms of
electrons (and electronic holes), note that in spite of the charge
$-2e$ (and $+2e$) of the $-1/2$ (and $+1/2$) Yang holons, these
quantum objects do not contribute to charge transport. The reason
is that they are the same quantum object as the corresponding
rotated $-1/2$ (and $+1/2$) Yang holons and thus refer to
localized electron doubly-occupied (and empty) sites for all
values of $U/t$. In contrast, for finite values of $U/t$ both the
$c$ pseudoparticles and composite $c\nu$ pseudoparticles are in
general not invariant under the electron - rotated-electron
unitary transformation and have a momentum dependent energy
dispersion, whose expression is given in Eqs.
(\ref{ep0c0})-(\ref{ep0sn}) of Appendix C. According to the
electron double-occupation studies of Ref. \cite{II}, these
quantum objects correspond to complex electron site distribution
configurations. Thus, both the associated chargeons (or
antichargeons) and $\nu$ $-1/2$ holons (or $+1/2$ holons)
contribute to charge transport. On the other hand, it is found in
Ref. \cite{II} that in the limit $U/t\rightarrow\infty$ the $\nu$
$-1/2$ holons (and $+1/2$ holons) contained in a $c\nu$
pseudoparticle correspond to localized electron doubly occupied
sites (and empty sites). This behavior is associated with the fact
that as $U/t\rightarrow\infty$ the $c\nu$ pseudoparticle becomes
the same quantum object as the rotated $c\nu$ pseudoparticle.
Thus, in spite of the charge $-2\nu\,e$ (and $+2\nu\,e$) of the
$\nu$ $-1/2$ holons (and $+1/2$ holons) which are part of the
$c\nu$ pseudoparticle, in the limit $U/t\rightarrow\infty$ such a
quantum object does not contribute to charge transport. This
effect is also associated with the fact that all $\eta$-spin holon
occupancy configurations become degenerated in that limit. In that
limit, the only charge carriers are the chargeons or antichargeons
associated with the $c$ pseudoparticles. As
$U/t\rightarrow\infty$, the energy dispersion of these quantum
objects becomes a free spinless-fermion spectrum \cite{Penc}.

Note that when the initial state is a ground state, the deviations
$\Delta L_{\alpha,\,\pm 1/2}$ and the pseudoparticle momentum
distribution function deviations $\Delta N_{c} (q)$ and $\Delta
N_{\alpha\nu} (q)$ which appear in Eq. (\ref{chargeNq}) for
$\alpha =c$ are relative to the ground-state values given in Eqs.
(\ref{Ns10})-(\ref{Ncnsn0}) and (\ref{NcGS})-(\ref{NsGS}) of
Appendix C and read,

\begin{equation}
\Delta L_{\alpha,\,\pm 1/2} = L_{\alpha,\,\pm 1/2} \, ;
\hspace{1cm} \Delta N_{c} (q)=N_{c} (q)-N^0_{c} (q) \, ;
\hspace{1cm} \Delta N_{\alpha\nu} (q)=N_{\alpha\nu} (q) \, .
\label{DNq}
\end{equation}

The charge carried by the holons was found here from analysis of
Eq. (\ref{chargeNq}). The charge carried by a $c$ pseudoparticle
and a composite $c\nu$ pseudoparticle can be obtained by other
methods. With the use of periodic boundary conditions, the charge
transport properties of the model can be studied by threading the
$N_a$-site ring by a flux \cite{Nuno}. That scheme leads to a
generalization of the Takahshi's thermodynamic equations
(\ref{Tapco1})-(\ref{Tapco3}), including the dependence on the
flux. From the use of that method one finds for the charge
transport in terms of electrons that the charge carried by a $c$
pseudoparticle is $-e$, whereas the charge carried by a composite
$c\nu$ pseudoparticle is $-2e\,\nu$. These results are consistent
with the charges found here for the holons.

Spin transport can also be described in terms of electrons or
electronic holes. For the electronic description the $N_c$ spinons
correspond to the $N_c$ rotated electrons of the singly occupied
sites. On the other hand, the electronic hole description of spin
transport corresponds to identifying the $N_c$ spinons with the
$N_c$ rotated-electronic holes which correspond to
rotated-electron singly occupied sites. The spin projection of a
electronic hole is the opposite of the corresponding electron.
However, we are reminded that the composite $s\nu$ pseudoparticles
contain an equal number $\nu$ of spin-down and spin-up spinons.
Thus, their number of spin-down and spin-up spinons remains
invariant under the change from the electron to the electronic
hole description of spin transport.

As for the transport of charge, quantum objects which have spin
but are invariant under the electron - rotated-electron unitary
transformation are localized and do not contribute to the
transport of spin. This is the case of the HL spinons for all
values of $U/t$ and of the composite $s\nu$ pseudoparticles for
$U/t\rightarrow\infty$. Furthermore, according to the results of
Ref. \cite{II}, composite $s\nu$ pseudoparticles with momentum
$q=\pm q_{s\nu}$ are localized for all values of $U/t$ and do not
contribute to the transport of spin. Importantly, for zero spin
density the momentum values reduce to $q=\pm q_{s\nu}=0$ for
$s\nu$ pseudoparticles belonging to branches such that $\nu >1$.
It follows that for zero spin density the $s\nu$ pseudoparticles
such that $\nu
>1$ are localized and do not contribute to the transport of
spin for all values of $U/t$. Thus, for zero spin density the only
quantum objects which contribute to the transport of spin are the
spinons contained in the two-spinon $s1$ pseudoparticles.

A similar situation occurs for the transport of charge at half
filling. For such a density the momentum values reduce to $q=\pm
q_{c\nu}=0$ for the $c\nu$ pseudoparticle branches. It follows
that the composite $c\nu$ pseudoparticles are localized and do not
contribute to the transport of charge for all values of $U/t$. For
values of excitation energy/frequency smaller than the
Mott-Hubbard gap \cite{Lieb}, the $c$ pseudoparticle band is full
and there is no transport of charge for the half-filling insulator
phase. For values of excitation energy/frequency larger than the
Mott-Hubbard gap, the only half-filling carriers of charge are the
chargeons or antichargeons contained in the $c$ pseudoparticles.
In summary, these results confirm the importance for the transport
of charge and spin of the transformation laws under the electron -
rotated-electron unitary rotation.

\section{SUMMARY AND CONCLUDING REMARKS}

In this paper we have shown that the rotated electrons play a
central role in the relation of the electrons to the exotic
quantum objects whose occupancy configurations describe all energy
eigenstates of the 1D Hubbard model. Such a clarification revealed
that as a result of the Hilbert-space electron - rotated-electron
unitary rotation, the $\eta$-spin $1/2$ holons, spin $1/2$
spinons, and $c$ pseudoparticles emerge. The energy eigenstates of
the 1D Hubbard model can be described in terms of occupancy
configurations of these three elementary quantum objects only, for
{\it all} finite values of energy and for the whole parameter
space of the model.

Our results also reveal the relation of the excitations associated
with Takahashi's ideal charge and spin strings of length
$\nu=1,2,...$, obtained by means of the Bethe-ansatz solution to
the $\pm 1/2$ holons and $\pm 1/2$ spinons, respectively. These
charge and spin excitations can be described in terms of occupancy
configurations of $2\nu$-holon composite $c\nu$ pseudoparticles
and $2\nu$-spinon composite $s\nu$ pseudoparticles, respectively.
In contrast to the holons (and spinons) which are part of
composite quantum objects, the $\pm 1/2$ Yang holons (and $\pm
1/2$ HL spinons) are invariant under the electron -
rotated-electron transformation for all values of $U/t$ and thus
have a localized character.

We found that the transformation laws under the electron -
rotated-electron unitary rotation of the quantum objects whose
occupancy configurations describe the energy eigenstates play a
crucial role in the transport of charge and spin. Indeed, quantum
objects which have charge (or spin), but are invariant under such
a unitary rotation are localized and do not contribute to the
transport of charge (or spin). Moreover, we found that for finite
values of $U/t$ and zero magnetization, the transport of charge
(or spin) is described by the charge $c$ pseudoparticle and $c\nu$
pseudoparticle quantum fluids (or the spin $s1$ pseudoparticle
quantum fluid). Within the electronic description of transport,
the elementary carriers of charge were found to be the charge $-e$
chargeon and the charge $-2e$ and $\eta$-spin projection $-1/2$
holon. Interestingly, the spin singlet and composite two-spinon
character of the $s1$ pseudoparticles reveals that the quantum
spin fluid described by these quantum objects is the 1D
realization of the two-dimensional resonating valence bond spin
fluid \cite{Anderson}.

Several authors described the low-energy excitations of the 1D
Hubbard model in terms of holons and spinons
\cite{spectral0,Anderson,PWA,Penc,hs,Tohyama,Carmelo97}. However,
often such holons and spinons correspond to different definitions.
While the pseudoparticle description of Ref. \cite{Carmelo97} is
valid for the LWSs, the $\alpha ,\beta$ hole representation used
for the non-LWSs in that reference only applies to towers whose
LWSs have no charge strings and no spin strings of length $\nu
>1$. Our study reveals that the concept of holon and spinon is not
limited to low energy. Indeed, it refers to {\it all} energy
scales of the model. In this paper we found that the charge
carriers associated with the distribution of $k's$ of the
Bethe-ansatz solution are the chargeons or antichargeons. However,
many authors identified the energy spectrum of the distribution of
$k's$ with the holon spectrum. Such an interpretation is behind
the value of charge obtained for the $-1/2$ holon and $+1/2$ holon
in Ref. \cite{hs} (called antiholon and holon in that reference),
which is half of the value found here. The chargeons and
antichargeons carry indeed charge $-e$ and $+e$, respectively, but
are not related to the $\eta$-spin $SU(2)$ representation, as
confirmed in Sec. III. Only the $-1/2$ holons and $+1/2$ holons of
charge $-2e$ and $+2e$, respectively, defined in this paper
correspond to the $\eta$-spin $SU(2)$ representation. For large
values of $U/t$ the $-1/2$ holons and $+1/2$ holons introduced
here become the dublons and holons, respectively, considered in
Ref. \cite{Tohyama}. Thus, for large values of $U/t$ our holon
definition is the same as in that reference.

The holon, spinon, and $c$ pseudoparticle description introduced
in this paper is useful for the evaluation of few-electron
spectral function expressions for finite values of energy. In this
paper we have not addressed the pseudoparticle problem in terms of
an operator algebra. However, the concepts introduced here are
used in Refs. \cite{V,spectral} in the construction of a suitable
operator algebra for the study of matrix elements of few-electron
operators between energy eigenstates. That algebra involves the
$c$ pseudofermion and composite $\alpha\nu$ pseudofermions, which
carry momentum $q+Q_c (q)/L$ and $q+Q_{\alpha\nu} (q)/L$,
respectively. Here the values of the momenta $Q_c (q)/L$ and
$Q_{\alpha\nu} (q)/L$ are determined by phase shifts. These
objects carry the same charge or spin and, in case of having a
composite character, have the same holon or spinon contents as the
correspondig $c$ pseudoparticle and composite $\alpha\nu$
pseudoparticles. They are related to the latter objects by a mere
momentum unitary transformation such that $q\rightarrow q+Q_c
(q)/L$ and $q\rightarrow q+Q_{\alpha\nu} (q)$. While the results
obtained in this paper play a major role in the studies of Refs.
\cite{V,spectral}, the expression of the problem in terms of
pseudofermion operators is most suitable for the evaluation of the
above-mentioned matrix elements. The motivation of the
investigations of both the present paper and Refs.
\cite{V,spectral} is the study of the unconventional finite-energy
spectral properties observed in low-dimensional materials
\cite{spectral0,Gweon}. Applications of the holon and spinon
description introduced in this paper and in Ref. \cite{V} to the
study of the one-electron removal spectral properties of the
organic conductor TTF-TCNQ are presented in Refs.
\cite{spectral0,spectral}.

\begin{acknowledgments}
We thank Katrina Hibberd, Lu\'{\i}s Miguel Martelo, Jo\~ao Lopes
dos Santos, and Pedro Sacramento for stimulating discussions. We
also thank the support of EU money Center of Excellence
ICA1-CT-2000-70029, K.P. thanks the support of OTKA grant T037451,
and J. M. R. thanks the financial support of FCT (Portugal) under
the fellowship No. SFRH/BPD/3636/2000.
\end{acknowledgments}
\appendix

\section{RELATION TO TAKAHASHI'S NOTATION}

By using his string hypothesis in the general Bethe-ansatz
equations for the model and taking logarithms, Takahashi arrived
to a system of coupled non-linear equations \cite{Takahashi}.
These equations introduce the quantum numbers $I_j$,
${J'}_{\alpha}^n$, and $J_{\alpha}^n$ whose occupancy
configurations describe the energy eigenstates. The occupied and
unoccupied values of these numbers are integers or half-odd
integers. The above-mentioned non-linear equations involve {\it
rapidities} ${\Lambda '}_{\alpha}^n$ and $\Lambda_{\alpha}^n$.
These are the real parts of the $N_a\rightarrow\infty$ ideal
charge, $\Lambda '$, and spin, $\Lambda $, strings of length $n$,
respectively.

It is convenient for our description to replace the above
Takahashi indices $\alpha$ and $n=1,2,...$ by $j$ and $\nu
=1,2,...$, respectively. In Appendix B we express the
pseudoparticle discrete momentum values in terms of the above
Takahashi's quantum numbers which within our notation read,

\begin{equation}
I_j^c\equiv I_j \, ; \hspace{1cm} I_j^{c\nu}\equiv {J'}_j^{\nu} \,
; \hspace{1cm} I_j^{s\nu}\equiv J_j^{\nu} \, . \label{I}
\end{equation}
By use of the notation relation given in this equation all
expressions introduced in Appendix B can be directly obtained from
the expressions provided in Ref. \cite{Takahashi}.

\section{THE PSEUDOPARTICLE DESCRIPTION}

The pseudoparticle description corresponds to associating the
quantum numbers

\begin{equation}
q_j = {2\pi\over L}\, I_j^c \,  ; \hspace{1cm} q_j = {2\pi\over
L}\, I_j^{c\nu} \,  ; \hspace{1cm} q_j = {2\pi\over L}\,
I_j^{s\nu}\, , \label{qj}
\end{equation}
such that

\begin{equation}
q_{j+1}-q_j = {2\pi\over L} \, , \label{separation}
\end{equation}
with the discrete momentum values of $c$ pseudoparticles, $c\nu$
pseudoparticles, and $s\nu$ pseudoparticles. From the properties
of Takahashi's equations and associated quantum numbers, we find
that the pseudoparticles obey a Pauli principle relative to the
momentum occupancies, {\it i.e.} one discrete momentum value $q_j$
can either be unoccupied or singly occupied by a pseudoparticle.
Within such a pseudoparticle description the set of $N_c$ and
$N_{\alpha\nu}$ occupied $q_j$ discrete momentum values correspond
to the number of pseudoparticles in the $c$ and $\alpha\nu$ bands,
respectively.

Since the energy eigenstates are uniquely defined by the
occupation numbers of the discrete momentum values $q_j$ of the
$c$ and $\alpha\nu$ pseudoparticle bands, it is useful to
introduce $c$ and $\alpha\nu$ momentum distribution functions $N_c
(q)$ and $N_{\alpha\nu} (q)$. These distributions read $N_c
(q_j)=1$ and $N_{\alpha\nu} (q_j)= 1$ for occupied values of the
discrete momentum values $q_j$ and $N_c (q_j)=0$ and
$N_{\alpha\nu} (q_j)= 0$ for unoccupied values of $q_j$. We also
introduce the following pseudoparticle-hole momentum distribution
functions,

\begin{equation}
N_c^h (q_j) \equiv 1 - N_c (q_j) \, ; \hspace{1cm} N_{\alpha\nu}^h
(q_j) \equiv 1 - N_{\alpha\nu} (q_j) \, . \label{NNh1}
\end{equation}
All LWSs of the $\eta$-spin and spin $SU(2)$ algebras are uniquely
defined by the set of infinite momentum distribution functions
$N_c (q)$ and $\{N_{\alpha\nu} (q)\}$ with $\alpha =c,s$ and $\nu
=1,2,...$.

Moreover, Takahashi's rapidity momentum $k_j$, charge rapidity
${\Lambda '}_{j}^{\nu}$, and spin rapidity  $\Lambda_{j}^{\nu}$
can be written as follows,

\begin{equation}
k(q_j)\equiv k_j \, ; \hspace{1cm} \Lambda_{c\nu} (q_j) \equiv
{\Lambda '}_j^{\nu } \, ; \hspace{1cm} \Lambda_{s\nu} (q_j) \equiv
\Lambda_j^{\nu} \, . \label{nlamb}
\end{equation}
The value of the rapidity momentum functional $k(q)$ and rapidity
functional $\Lambda_{c\nu} (q)$ is defined by Eqs.
(\ref{Tapco1})-(\ref{Tapco3}) where the function
$\Theta_{\nu,\,\nu'}(x)$ is given by,

\begin{eqnarray}
\Theta_{\nu,\,\nu'}(x) & = & \Theta_{\nu',\nu}(x) = \delta_{\nu
,\nu'}\Bigl\{2\arctan\Bigl({x\over 2\nu}\Bigl) + \sum_{l=1}^{\nu
-1}4\arctan\Bigl({x\over 2l}\Bigl)\Bigr\}
\nonumber \\
& + & (1-\delta_{\nu ,\nu'})\Bigl\{ 2\arctan\Bigl({x\over |\nu
-\nu'|}\Bigl) + 2\arctan\Bigl({x\over \nu + \nu'}\Bigl) +
\sum_{l=1}^{{\nu +\nu' -|\nu -\nu'|\over 2}
-1}4\arctan\Bigl({x\over |\nu -\nu'|+2l}\Bigl)\Bigr\} \, .
\label{Theta}
\end{eqnarray}

The discrete momentum values $q_j$ of the $\alpha\nu$ (or $c$)
pseudoparticles are such that $j = 1,2,...,N^*_{\alpha\nu}$ (or
$j=1,2,...,N_a$), where

\begin{equation}
N^*_{\alpha\nu}=N_{\alpha\nu}+ N^h_{\alpha\nu} \, , \label{N*}
\end{equation}
(or $N_a=N_c+N_c^h$) and $N^h_{\alpha\nu}$ (or $N^h_c$) is the
number of unoccupied momentum values in the $\alpha\nu$ (or $c$)
pseudoparticle band. Both the numbers $N^*_{\alpha\nu}$ and the
associate numbers $N^h_{\alpha\nu}$ are dependent and fully
determined by the values of the set of numbers $N_c$ and
$\{N_{\alpha\nu'}\}$, where $\alpha =c,s$ and $\nu' =\nu +1,\nu
+2,\nu +3,...$, of occupied $c$ and $\alpha\nu'$ pseudoparticle
momentum values, respectively. The numbers $N^h_{\alpha\nu}$ of
$\alpha\nu$ pseudoparticles and $N^h_c$ of $c$ pseudoparticle
holes are given by

\begin{equation}
N^h_{\alpha\nu} = 2\,S_{\alpha} + 2\sum_{\nu'=\nu +1}^{\infty}
(\nu' -\nu) N_{\alpha\nu'} = L_{\alpha} + 2\sum_{\nu'=\nu
+1}^{\infty} (\nu' -\nu) N_{\alpha\nu'} \, , \label{Nhag}
\end{equation}
and

\begin{equation}
N^h_c = N_a -N_c \, , \label{Nhc}
\end{equation}
respectively. Here $L_c$ and $L_s$ are the numbers of Yang holons
and HL spinons, respectively, and the $\eta$-spin and spin values
$S_c=L_c/2$ and $S_s=L_s/2$, respectively, are also determined by
the values of the pseudoparticle numbers and are given in Eqs.
(\ref{Scco}) and (\ref{Ssco}). The pseudoparticle and
pseudoparticle-hole momentum distribution functions are such that

\begin{equation}
\sum_{j=1}^{N_a}\, N_c (q_j) = N_c \, ; \hspace{1cm}
\sum_{j=1}^{N_a}\, N_c^h (q_j) = N_a -N_c \, , \label{NNhc}
\end{equation}

\begin{equation}
\sum_{j=1}^{N_{\alpha\nu}^*}\, N_{\alpha\nu} (q_j) = N_{\alpha\nu}
\, ; \hspace{1cm} \sum_{j=1}^{N_{\alpha\nu}^*}\, N_{\alpha\nu}^h
(q_j) = N_{\alpha\nu}^h \, . \label{NNhcg}
\end{equation}
From the combination of Eqs. (\ref{Scco}), (\ref{Ssco}),
(\ref{Nhag}), (\ref{NNhc}), and (\ref{NNhcg}) the number of $c\nu$
and $s\nu$ pseudoparticle holes (\ref{Nhag}) can be rewritten as
follows,

\begin{equation}
N^h_{c\nu} = N_a -N_c - \sum_{\nu'=1}^{\infty} \Bigl(\nu + \nu' -
\vert\nu - \nu'\vert\Bigl) N_{c\nu'} \, ; \hspace{0.5cm}
N^h_{s\nu} = N_c - \sum_{\nu'=1}^{\infty} \Bigl(\nu + \nu' -
\vert\nu - \nu'\vert\Bigl) N_{s\nu'} \, . \label{Nhag2}
\end{equation}
For the ground-state CPHS ensemble subspace of densities in the
ranges $0\leq n\leq 1$ and $0\leq m\leq n$ considered in Appendix
C, the expression of the number $N^*_{\alpha\nu}$ given in Eqs.
(\ref{N*}), (\ref{Nhag}), and (\ref{Nhag2}) simplifies to,

\begin{equation}
N^*_{c\nu}=(N_a -N) \, ; \hspace{1cm} N^*_{s1}=N_{\uparrow} \, ;
\hspace{1cm} N^*_{s\nu}=(N_{\uparrow} -N_{\downarrow}) \, ,
\hspace{0.3cm} \nu > 1 \, , \label{N*csnu}
\end{equation}
whereas $N^*_c$ is given by $N^*_c=N_a$ for all energy
eigenstates.

The $I_j^{\alpha\nu}$ numbers of Eq. (\ref{qj}) are integers
(half-odd integers), if $N^*_{\alpha\nu}$ is odd (even) and thus
the $q_j$ discrete momentum values obey Eq. (\ref{separation}),
are distributed symmetrically about zero, and are such that

\begin{equation}
\vert q_j\vert \leq q_{\alpha\nu} \, , \label{rangeqj}
\end{equation}
where

\begin{equation}
q_{\alpha\nu} = {\pi\over L}[N^*_{\alpha\nu}-1] \approx {\pi
N^*_{\alpha\nu}\over L} \, . \label{qag}
\end{equation}
On the other hand, the $I_j^c$ numbers of Eq. (\ref{qj}) are
integers (half-odd integers), if ${N_a\over 2} - \sum_{\alpha
=c,s}\sum_{\nu = 1}^{\infty}N_{\alpha\nu}$ is odd (even) and the
momentum values $q_j$ obey Eq. (\ref{separation}) and are such
that

\begin{equation}
q_c^{-} \leq q_j \leq q_c^{+} \, , \label{rangeqjc}
\end{equation}
where

\begin{equation}
q_c^{+} = - q_c^{-}= {\pi\over a}\bigl[1-{1\over N_a}\bigr] \, ,
\label{qcev}
\end{equation}
for ${N_a\over 2} - \sum_{\alpha =c,s}\sum_{\nu =
1}^{\infty}N_{\alpha\nu}$ even and

\begin{equation}
q_c^{+} = {\pi\over a} \, ; \hspace{1cm} q_c^{-} = - {\pi\over a}
\bigl[1-{2\over N_a}\bigr] \, , \label{qcodd}
\end{equation}
for ${N_a\over 2} - \sum_{\alpha =c,s}\sum_{\nu =
1}^{\infty}N_{\alpha\nu}$ odd.

\section{THE GROUND-STATE DISTRIBUTIONS AND PSEUDOPARTICLE ENERGY BANDS}

In this Appendix we provide both the ground-state pseudoparticle
momentum distribution functions and the expressions which define
the pseudoparticle energy bands. The specific $c$ and $\alpha\nu$
pseudoparticle momentum distribution functions which describe a
ground state corresponding to values of the densities in the
ranges $0\leq n\leq 1/a$ and $0\leq m\leq n$ read
\cite{Carmelo97},

\begin{equation}
N^{0}_c (q) = \Theta\Bigl(q_{Fc}^{+} - q \Bigl) \, ,
\hspace{0.5cm} 0\leq q\leq q_c^{+} \, ; \hspace{1cm} N^{0}_c (q) =
\Theta\Bigl(q - q_{Fc}^{-}\Bigl) \, , \hspace{0.5cm} q_c^{-}\leq
q\leq 0 \, ; \label{Nc0}
\end{equation}

\begin{equation}
N^{0}_{s1} (q) = \Theta\Bigl(q_{Fs1} - q \Bigl) \, ,
\hspace{0.5cm} 0\leq q\leq q_{s1} \, ; \hspace{1cm} N^{0}_{s1} (q)
= \Theta\Bigl(q + q_{Fs1}\Bigl) \, , \hspace{0.5cm} -q_{s1}\leq
q\leq 0 \, ; \label{Ns10}
\end{equation}
and

\begin{equation}
N^{0}_{\alpha\nu}(q) = 0 \, ; \hspace{0.5cm} -q_{\alpha\nu}\leq
q\leq q_{\alpha\nu} \, ; \hspace{0.5cm} \alpha = c,\,s  \, ;
\hspace{0.5cm} \nu \geq 1 + \delta_{\alpha ,\,s} \, .
\label{Ncnsn0}
\end{equation}
In these equations, the momentum limiting values $q_{s1}$,
$q_{c\nu}$, and $q_{s\nu}$ for $\nu>1$ are given in Eq.
(\ref{qag}) of Appendix B and $q_{c}^{\pm}$ is defined in Eqs.
(\ref{rangeqjc})-(\ref{qcodd}) of the same Appendix. Moreover, the
expressions of the $c$ pseudoparticle ground-state {\it Fermi}
momentum values are given by,

\begin{equation}
q_{Fc}^{\pm} = \pm q_{Fc} \, , \label{qFcsipm}
\end{equation}
for both $N$ and ${N_a\over 2} - \sum_{\alpha =c,s}\sum_{\nu =
1}^{\infty}N_{\alpha\nu}$ even or odd and by

\begin{equation}
q_{Fc}^{+} = q_{Fc} + {\pi\over L}\, ; \hspace{1cm} q_{Fc}^{-} =
-q_{Fc} + {\pi\over L} \, , \label{qFcan1}
\end{equation}
or by

\begin{equation}
q_{Fc}^{+} = q_{Fc} - {\pi\over L}\, ; \hspace{1cm} q_{Fc}^{-} =
-q_{Fc} - {\pi\over L} \, , \label{qFcan2}
\end{equation}
for ${N_a\over 2} - \sum_{\alpha =c,s}\sum_{\nu =
1}^{\infty}N_{\alpha\nu}$ even or odd and $N$ odd or even,
respectively, where

\begin{equation}
q_{Fc} = {\pi\over a} \bigl[n-{1\over N_a}\bigr] = 2k_F + O(1/L)
\, . \label{qFcsi}
\end{equation}

Furthermore, the $s1$ pseudoparticle ground-state {\it Fermi}
momentum values read

\begin{equation}
q_{Fs1}^{\pm} = \pm q_{Fs1} \, , \label{qFs1pm}
\end{equation}
for both $N_{\uparrow}$ and $N_{\downarrow}$ even or odd and

\begin{equation}
q_{Fs1}^{+} = q_{Fs1} + {\pi\over L} \, ; \hspace{1cm} q_{Fs1}^{-}
= -q_{Fs1} + {\pi\over L} \, , \label{qs1odda}
\end{equation}
or

\begin{equation}
q_{Fs1}^{+} = q_{Fs1} - {\pi\over L} \, ; \hspace{1cm} q_{Fs1}^{-}
= -q_{Fs1} - {\pi\over L} \, , \label{qs1oddb}
\end{equation}
for $N_{\uparrow}$ even or odd and $N_{\downarrow}$ odd or even,
respectively, where

\begin{equation}
q_{Fs1} = {\pi\over a} \bigl[n_{\downarrow}-{1\over N_a}\bigr] =
k_{F\downarrow} + O(1/L) \, . \label{qss1}
\end{equation}

In the particular case of the ground-state momentum function
distributions (\ref{Nc0})-(\ref{Ncnsn0}), the momentum limiting
values given by Eqs. ({\ref{qag})-(\ref{qcodd}) of Appendix B,
including $1/L$ contributions, simplify to,

\begin{equation}
q_c = {\pi\over a}\bigl[1-{1\over N_a}\bigr] = \pi + O(1/L) \, ;
\hspace{1cm} q_{s1} = {\pi\over L}[N_{\uparrow} -1] =
k_{F\uparrow} + O(1/L)\, , \label{qcs1}
\end{equation}
and

\begin{equation}
q_{c\nu} = {\pi\over L}[N_a -N -1] = (\pi - 2k_F) + O(1/L) \, ,
\label{qcnpi2k}
\end{equation}

\begin{equation}
q_{s\nu} = {\pi\over L}[N_{\uparrow} - N_{\downarrow} -1] =
(k_{F\uparrow} - k_{F\downarrow}) + O(1/L) \, ; \hspace{0.5cm} \nu
> 1 \, . \label{qsnkk}
\end{equation}

The pseudoparticle energy bands read \cite{Carmelo97},

\begin{equation}
\epsilon_c (q) = a\,\int_Q^{k^{0}(q)} dk\, 2t\eta_c (k) \, ,
\label{ep0c0}
\end{equation}

\begin{equation}
\epsilon_{s1} (q) = \int_{\infty}^{\Lambda^{0}_{s1} (q)}
d\Lambda\, 2t\eta_{s,\,1}(\Lambda) + 2\mu_0 h \, , \label{ep0s1}
\end{equation}

\begin{equation}
\epsilon_{c\nu}^0(q) = \int_{\infty}^{\Lambda^{(0)}_{c\nu}(q)}
d\Lambda\, 2t\eta_{c\nu}(\Lambda) \, , \label{ep0cn}
\end{equation}

\begin{equation}
\epsilon_{s\nu}^0 (q) = \int_{\infty}^{\Lambda^{0}_{s\nu} (q)}
d\Lambda\, 2t\eta_{s\nu}(\Lambda) \, , \label{ep0sn}
\end{equation}
Here $k^{0}(q)$, $\Lambda^{0}_{s1} (q)$,
$\Lambda^{(0)}_{c\nu}(q)$, and $\Lambda^{0}_{s\nu} (q)$ are the
values specific to the ground-state. These values are obtained by
solution of the integral equations (\ref{Tapco1})-(\ref{Tapco3})
in the particular case when the pseudoparticle momentum
distribution functions are given by Eqs.
(\ref{Nc0})-(\ref{Ncnsn0}). Moreover, the functions $2t\eta_c
(k)$, $2t\eta_{c\nu}(\Lambda)$, and $2t\eta_{s\nu}(\Lambda)$ are
defined by the following integral equations,

\begin{equation}
2t\eta_c (k) = 2t\sin (k\,a) + {4t\,\cos (k\,a)\over \pi}\,
\int_{-B}^{B}d\Lambda\, {U \over U^2 + (4t)^2(\Lambda -\sin
(k\,a))^2}\,2t\eta_{s1}(\Lambda) \, , \label{etac0}
\end{equation}

\begin{equation}
2t\eta_{c\nu}(\Lambda) = - 4t Re\,\Bigl({4t\,\Lambda - i\nu U\over
\sqrt{(4t)^2 - (4t\,\Lambda - i\nu U)^2}}\Bigl) + {4t\over
\pi}\int_{-Q}^{Q}dk\, {\nu U \over (\nu U)^2+(4t)^2(\sin (k\,a)
-\Lambda)^2}\,2t\eta_{c}(k) \, ; \hspace{0.5cm} \nu\neq 0 \, ,
\label{etacn}
\end{equation}
and

\begin{equation}
2t\eta_{s\nu}(\Lambda) = {4t\over \pi} \int_{-Q}^{Q}dk\, {\nu
U\over (\nu U)^2+(4t)^2(\sin (k\,a) - \Lambda)^2}\,2t\eta_{c}(k) -
{2t\over \pi U}\int_{-B}^{B}d\Lambda'\,
\Theta^{[1]}_{1,\,\nu}\Bigl({4t(\Lambda'- \Lambda)\over U}
\Bigl)\,2t\eta_{s1}(\Lambda') \, . \label{etasn}
\end{equation}
Here,

\begin{eqnarray}
\Theta^{[1]}_{\nu,\,\nu'}(x) & = & {d\Theta_{\nu,\,\nu'}(x)\over
dx} = \delta_{\nu ,\nu'}\{{1\over \nu[1+({x\over 2\nu})^2]}+
\sum_{l=1}^{\nu -1}{2\over l[1+({x\over 2l})^2]}\} +
(1-\delta_{\nu ,\nu'})\{{2\over |\nu-\nu'|[1+({x\over
|\nu-\nu'|})^2]}
\nonumber \\
& + &  {2\over (\nu+\nu')[1+({x\over \nu+\nu'})^2]} +
\sum_{l=1}^{{\nu+\nu'-|\nu-\nu'|\over 2} -1}{4\over
(|\nu-\nu'|+2l)[1+({x\over |\nu-\nu'|+2l})^2]}\} \, , \label{The1}
\end{eqnarray}
is the $x$ derivative of the function given in Eq. (\ref{Theta})
of Appendix B and the parameters $Q$ and $B$ are defined by,

\begin{equation}
Q = k^{0}(2k_F) \, , \hspace{1cm} B =
\Lambda^{0}_{s1}(k_{F\downarrow}) \, . \label{QB}
\end{equation}

The pseudoparticle momentum distribution functions
(\ref{Nc0})-(\ref{Ncnsn0}) correspond to the ground state. We note
that the pseudoparticle occupancies of such a state were studied
in Refs. \cite{Carmelo97,Carmelo92}. Combining the pseudoparticle
language of these references with the holon and spinon description
introduced in this paper, we find that the ground state is
characterized by the following values for the pseudoparticle, $\pm
1/2$ holon, and $\pm 1/2$ spinon numbers,

\begin{equation}
M^0_{c,\,-1/2} = L^0_{c,\,-1/2} = 0 \, ; \hspace{0.5cm}
M^0_{c,\,+1/2} = L^0_{c,\,+1/2} = N_a - N \, ; \hspace{0.5cm}
N^0_c = N \, ; \hspace{0.5cm} N^0_{c\nu} = 0 \, , \label{NcGS}
\end{equation}
in the charge sector and

\begin{equation}
M^0_{s,\,-1/2} = N^0_{s1} = N_{\downarrow} \, ; \hspace{0.5cm}
M^0_{s,\,+1/2} = N_{\uparrow} \, ; \hspace{0.5cm} L^0_{s,\,-1/2} =
N^0_{s\nu} = 0 \, , \hspace{0.25cm} \nu\geq 2 ; \hspace{0.5cm}
L^0_{s,\,+1/2} = N_{\uparrow}-N_{\downarrow} \, , \label{NsGS}
\end{equation}
in the spin sector.



\begin{references}
\bibitem[1]{Lieb}
        Elliott H. Lieb and F. Y. Wu, Phys. Rev. Lett. {\bf 20},
        1445 (1968); P. B. Ramos and M. J. Martins , J. Phys. A {\bf 30}, L195 (1997);
        M. J. Martins and P.B. Ramos, Nucl. Phys. B {\bf 522}, 413 (1998).
\bibitem[2]{Takahashi}
        M. Takahashi, Prog. Theor. Phys. {\bf 47}, 69 (1972).
\bibitem[3]{spectral0}
        M. Sing, U. Schwingenschl\"ogl, R. Claessen, P.
        Blaha, J. M. P. Carmelo, L. M. Martelo, P. D. Sacramento, M.
        Dressel, and C. S. Jacobsen, Phys. Rev. B {\bf 68}, 125111 (2003);
        R. Claessen, M. Sing, U. Schwingenschl\"ogl, P. Blaha,
        M. Dressel, and C. S. Jacobsen, Phys. Rev. Lett. {\bf 88}, 096402
        (2002).
\bibitem[4]{Gweon}
        G.-H. Gweon, J. D. Denlinger, J. W. Allen, R. Claessen, C. G. Olson, H. Hochst,
        J. Marcus, C. Schlenker, and L. F. Schneemeyer, J. Electron Spectrosc. Relat.
        Phenom. {\bf 117-118}, 481 (2001).
\bibitem[5]{Granath}
        M. Granath, V. Oganesyan, D. Orgad, and S. A. Kivelson, Phys. Rev. B {\bf 65},
        184501 (2002); E. W. Carlson, D. Orgad, S. A. Kivelson, and  V. J. Emery,
        {\it ibid.} {\bf 62}, 3422 (2000).
\bibitem[6]{Zaanen}
        J. Zaanen, O. Y. Osman, H. V. Kruis, Z. Nussinov, and J.
        Tworzydlo, Philos. Magaz. B {\bf 81}, 1485 (2001).
\bibitem[7]{Antonio}
        A. L. Chernyshev, S. R. White, and A. H. Castro Neto, Phys.
        Rev. B {\bf 65}, 214527 (2002).
\bibitem[8]{HL}
        O. J. Heilmann and E. H. Lieb, Ann. N. Y.
        Acad. Sci. {\bf 172}, 583 (1971);
        E. H. Lieb, Phys. Rev. Lett. {\bf 62},
        1201 (1989).
\bibitem[9]{Yang89}
        C. N. Yang, Phys. Rev. Lett. {\bf 63}, 2144 (1989).
\bibitem[10]{II}
        J. M. P. Carmelo and P. D. Sacramento, Phys. Rev. B {\bf 68}, 085104
        (2003).
\bibitem[11]{Essler}
        Fabian H. L. Essler, Vladimir E. Korepin, and
        Kareljan Schoutens, Phys. Rev. Lett. {\bf 67}, 3848 (1991).
\bibitem[12]{Korepin}
        Vladimir E. Korepin, in press in Phys. Rev. Lett. (2004) [cond-mat/0311056].
\bibitem[13]{Anderson}
        P. W. Anderson, in
        {\it The Theory of Superconductivity in the High-$T_c$
        Cuprates} (Princeton University Press, Princeton, NJ, 1997).
\bibitem[14]{PWA}
        J. C. Talstra, S. P. Strong, and P. W. Anderson, Phys. Rev. Lett. {\bf 74},
        5256 (1995); J. C. Talstra and S. P. Strong, Phys. Rev. B {\bf 56}, 6094
        (1997).
\bibitem[15]{Penc}
        Karlo Penc, Karen Hallberg, Fr\'ed\'eric Mila, and Hiroyuki Shiba,
        Phys. Rev. Lett. {\bf 77}, 1390 (1996).
\bibitem[16]{hs}
        F. H. L. Essler and V. E. Korepin, Phys. Rev. Lett. {\bf 72},
        908 (1994).
\bibitem[17]{Tohyama}
        Y. Mizuno, K. Tsutsui, T. Tohyama, and S. Maekawa, Phys. Rev.
        B {\bf 62}, R4769 (2000).
\bibitem[18]{Carmelo97}
        J. M. P. Carmelo and N. M. R. Peres, Phys. Rev. B
        {\bf 56}, 3717 (1997).
\bibitem[19]{Carmelo92}
        J. M. P. Carmelo, P. Horsch, and A. A. Ovchinnikov,
        Phys. Rev. B {\bf 45}, 7899 (1992).
\bibitem[20]{Harris}
        A. Brooks Harris and Robert V. Lange, Phys. Rev. {\bf 157},
        295 (1967); A. H. MacDonald, S. M. Girvin, and D. Yoshioka,
        Phys. Rev. B {\bf 37}, 9753 (1988).
\bibitem[21]{review}
        R. M. Barnett {\it et al.}, Phys. Rev. D {\bf 50}, 1173 (1994).
\bibitem[22]{V}
        J. M. P. Carmelo and K. Penc, cond-mat/0311075.
\bibitem[23]{spectral}
        J. M. P. Carmelo, K. Penc, L. M. Martelo, P. D. Sacramento,
        J. M. B. Lopes dos Santos, R. Claessen, M. Sing, and
        U. Schwingenschl\"ogl, cond-mat/0307602.
\bibitem[24]{Taka71}
        Minoru Takahashi, Prog. Theor. Phys. {\bf 46}, 401 (1971).
\bibitem[25]{Nuno}
        N. M. R. Peres, J. M. P. Carmelo, D. K. Campbell, and A. Sandvik,
        Z. Phys. B {\bf 103}, 217 (1997).
\end{references}
\end{document}